%% file: paper.tex
\documentclass[10pt,journal,compsoc]{IEEEtran}

\ifCLASSOPTIONcompsoc
  \usepackage[nocompress]{cite}
\else
  \usepackage{cite}
\fi
\ifCLASSINFOpdf
\else
\fi

\usepackage{algorithm}
\usepackage{algpseudocode}

\usepackage{array}

\usepackage{graphicx}
\usepackage{balance}  

\usepackage{amsmath}
\usepackage{url}
\usepackage{xspace}
\usepackage{subfigure}
\usepackage{color}
\usepackage{textcomp}
\usepackage{listings}
\usepackage{booktabs}
\begin{document}

\newcommand{\db}{DiNoDB\xspace}
\newcommand{\nodb}{PostgresRaw\xspace}
\newcommand{\plugin}{DiNoDB I/O decorators\xspace}
\newcommand{\abox}[1] {\vspace*{0.2cm} \noindent \hspace*{0cm}\fbox{ \begin{minipage}{0.95\columnwidth}
{#1}
\end{minipage}
}\\}
\newcommand{\omitit}[1]{}

%
\title{\db: an Interactive-speed Query Engine for Ad-hoc Queries on Temporary Data}
%
%
%
%

\author{Yongchao~Tian, Ioannis~Alagiannis, Erietta~Liarou, Anastasia~Ailamaki, Pietro~Michiardi, Marko~Vukoli{\'c}

\IEEEcompsocitemizethanks{\IEEEcompsocthanksitem Y. Tian and P. Michiardi are with EURECOM, F-06160 Biot, France. \protect\\
E-mail: (yongchao.tian, pietro.michiardi)@eurecom.fr
\IEEEcompsocthanksitem I. Alagiannis, E. Liarou and A. Ailamaki are with {\'E}cole Polytechnique F{\'e}d{\'e}rale de Lausanne, CH-1015 Lausanne, Switzerland.
E-mail: (ioannis.alagiannis, erietta.liarou, anastasia.ailamaki)@epfl.ch
\IEEEcompsocthanksitem M. Vukoli{\'c} is with IBM Research - Zurich, CH-8803 R{\"u}schlikon, Switzerland. 
E-mail: mvu@zurich.ibm.com
.}
\thanks{Manuscript received April 19, 2005; revised September 17, 2014.}}

%
%

\markboth{Journal of \LaTeX\ Class Files,~Vol.~13, No.~9, September~2014}%
{Shell \MakeLowercase{\textit{et al.}}: Bare Demo of IEEEtran.cls for Computer Society Journals}
%



\IEEEtitleabstractindextext{%
\begin{abstract}
\input{00-abstract}
\end{abstract}

}

\maketitle

\input{01-intro}

\input{02-applications}

\input{03-dinodb}

\input{04-experiments}

\input{05-related}

\input{07-conclusions}

\IEEEdisplaynontitleabstractindextext

%
\IEEEpeerreviewmaketitle

\ifCLASSOPTIONcompsoc
\else
\fi


\ifCLASSOPTIONcaptionsoff
  \newpage
\fi



%
{
\bibliographystyle{IEEEtran}
\bibliography{paper}

}

%






\end{document}

%% file: 00-abstract.tex
As data sets grow in size, analytics applications struggle to get instant insight into large datasets. Modern applications involve heavy \emph{batch processing} jobs over large volumes of data and at the same time require efficient \emph{ad-hoc interactive analytics} on temporary data. 
Existing solutions, however, typically focus on one of these two aspects, largely ignoring the need for synergy between the two. Consequently, interactive queries need to re-iterate costly passes through the entire dataset (e.g., data loading) that may provide meaningful return on investment only when data is queried over a long period of time.  
In this paper, we propose \db, an interactive-speed query engine for ad-hoc queries on temporary data. \db avoids the expensive loading and transformation phase that characterizes both traditional RDBMSs and current interactive analytics solutions. It is tailored to modern workflows found in machine learning and data exploration use cases, which often involve iterations of cycles of batch and interactive analytics on data that is typically useful for a narrow processing window. The key innovation of \db is to piggyback on the batch processing phase the creation of metadata that \db exploits to expedite the interactive queries.
Our experimental analysis
demonstrates that \db achieves very good performance for a wide range of ad-hoc queries compared to alternatives 
.

%% file: 01-intro.tex
\IEEEraisesectionheading{\section{Introduction}\label{sec:intro}}

\IEEEPARstart{I}n recent years, modern large-scale data analysis systems have flourished. For example, systems such as Hadoop and Spark~\cite{hadoop,spark-project} focus on issues related to fault-tolerance and expose a simple yet elegant parallel programming model that hides the complexities of synchronization. Moreover, the batch-oriented nature of such systems has been complemented by additional components (e.g., Storm and Spark streaming \cite{storm,spark_streaming}) that offer (near) real-time analytics on data streams. The communion of these approaches is now commonly known as the ``Lambda Architecture'' (LA)~\cite{lambda}. 
In fact, LA is split into three layers, i) the \emph{batch layer} (based on e.g., Hadoop/Spark) for managing and pre-processing append-only raw data, ii) the \emph{speed layer} (e.g., Storm/Spark streaming) tailored to analytics on recent data while achieving low latency using fast and incremental algorithms, and iii) the \emph{serving layer} (e.g., Hive~\cite{Hive}, SparkSQL~\cite{spark-project}, Impala~\cite{impala_cidr2015}) that exposes the batch views to support ad-hoc queries written in SQL, with low latency.

The problem with such 
existing large scale analytics systems 
is twofold. 
First, combining components (layers) from different stacks, though desirable, 
raises performance issues and is sometimes not even possible in practice.
For example, companies who have expertise 
in, e.g., Hadoop and traditional SQL-based (distributed) RDBMSs, 
would arguably like to leverage this expertise and use Hadoop as the batch processing layer 
and RDBMSs in the serving layer. 
However, this approach requires an expensive transform/load phase to, 
e.g., move data from Hadoop's HDFS and load it into a RDBMSs \cite{LeFevre:2014:MSU:2588555.2588568}, which might be impossible to
amortize, in particular in scenarios with a narrow processing window, \textit{i.e.,} when working on \emph{temporary data}.

Second, although many SQL-on-Hadoop systems emerged recently, they are not well designed for (short-lived) ad-hoc queries, especially when the data remains in its native, uncompressed, format such as text-based CSV files.
To achieve high performance, these systems ~\cite{Floratou:2014:SFC:2732977.2733002} prefer to convert data into their specific column-based data format, e.g., ORC~\cite{orc} and Parquet~\cite{parquet}. This works perfectly when both data and analytic queries (that is, the full workload) are in their final production stage. Namely, these self-describing, optimized data formats play an increasing role in modern data analytics, and this especially becomes true once data has been cleansed, queries have been well designed, and analytics algorithms have been tuned. 
However, when users perform data exploration tasks and algorithm tuning, that is when the data is \emph{temporary}, the original data format typically remains unchanged --- in this case, premature data format optimization is typically avoided, and simple text-based formats such as CSV and JSON files are preferred.
In this case, current integrated data analytics systems can under-perform. Notably, they often fail to leverage decades old techniques for optimizing the performance of (distributed) RDBMSs, \emph{e.g.,} indexing, that is usually not supported.

In summary, contemporary data scientists face a wide variety of competing approaches targeting the \emph{batch} 
and the \emph{serving} layer. Nevertheless, we believe that these approaches often have overly strict focus, in many cases ignoring one another, thus failing to explore potential benefits from learning from each other.

In this paper, we propose \db, an interactive-speed query engine that addresses the above issues. 
Our approach is based on a seamless integration 
of batch processing systems (e.g., Hadoop MapReduce and Apache Spark) 
with a distributed, fault-tolerant and scalable interactive query engine for \emph{in-situ} analytics on \emph{temporary} data. 
\db integrates the batch processing with the serving layer, by extending the ubiquitous Hadoop I/O API using \emph{\db I/O decorators}. This mechanism is used to create, as an additional output of batch processing, a wide range of \emph{metadata}, \textit{i.e.,} auxiliary data structures such as positional maps and vertical indexes, that \db uses to speed-up the interactive data analysis of \emph{temporary} data files for data exploration and algorithm tuning.
Our solution effectively brings together the batch processing and the serving layer for big data workflows, while avoiding any loading and data (re)formatting costs. While, clearly, no data analytics solution can fit all Big Data use cases, when it comes to ad-hoc interactive queries with a narrow processing window, \db outperforms state-of-the-art distributed query engines, such as Hive, Stado, SparkSQL and Impala.

In summary, our main contributions in this paper include:
\begin{itemize}

\item The design of \db, a distributed interactive query engine.
\db leverages modern multi-core architectures and provides efficient, distributed, fault-tolerant and scalable in-situ SQL-based querying capabilities for temporary data. \db (Distributed NoDB) is the first distributed and scalable instantiation of the NoDB paradigm  
\cite{Alagiannis:2012:NEQ:2213836.2213864}, which was previously instantiated only in centralized systems. 

\item Proposal and implementation of the \db I/O decorators approach 
to interfacing batch processing and interactive query serving engines in a data analytics system. 
\db I/O decorators generate, as a result of the batch processing phase, metadata that aims to facilitate and expedite subsequent interactive queries.

\item Detailed performance evaluation and comparative analysis of \db versus
state-of-the-art systems including Hive, Stado, SparkSQL and Impala.

\end{itemize}

The rest of the paper is organized as follows. In Section~\ref{sec:usecases}, we further motivate our approach and the need for a system such as \db. In Section~\ref{sec:architecture}, we describe the architecture of \db.  In Section~\ref{sec:eval} we give our experimental results based on both synthetic and real-life datasets. Section~\ref{sec:relwork} overviews related work. Section~\ref{sec:conclusion} concludes the paper.

%% file: 02-applications.tex
\section{Applications and use cases}
\label{sec:usecases}

In this section, we overview some of the contemporary uses cases 
which span both batch processing and interactive analytics in the data analytics flows. These use cases include machine learning (Section~\ref{sec:ml}) and data exploration (Section~\ref{sec:de}). 
For each of these use cases we discuss: i) how better communication between the batch processing and the serving layer that \db brings may help, and ii) the applicability of our \emph{temporary data} analytics approach.

\subsection{Machine learning}
\label{sec:ml}

\begin{figure}[t!]
  \center
  \includegraphics[width=1.0\columnwidth]{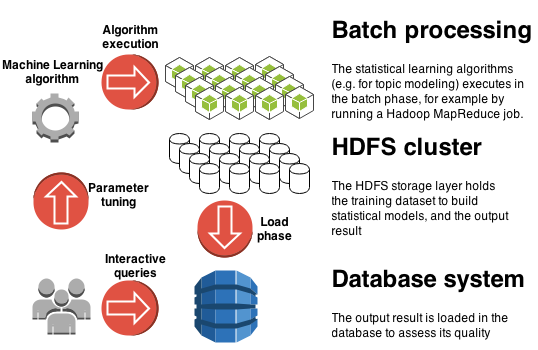}
  \caption{Machine learning use case.}
  \label{fig:ml}
\end{figure}

In the first use case -- which we evaluate in Section~\ref{sec:eval} -- we take the perspective of a user (e.g., a data scientist) focusing on a complex data clustering problem. Specifically, we consider the task of learning \emph{topic models}~\cite{lda}, which amounts to automatically and jointly clustering words into ``topics'',  and documents into mixtures of topics. Simply stated, a topic model is a hierarchical Bayesian model that associates with each document a probability distribution over ``topics'', which are in turn distributions over words. Thus, the output of a topic modeling data analysis can be thought of as a (possibly very large) matrix of probabilities: each row represents a document, each column a topic, and the value of a cell indicates the probability for a document to cover a particular topic.

In such a scenario, depicted in Figure~\ref{fig:ml}, the user typically faces the following issues: i) topic modeling algorithms (e.g.,  Collapsed Variational Bayes (CVB) \cite{cvb}) require parameter tuning, such as selecting an appropriate number of topics, the number of unique features to consider, distribution smoothing factors, and many more; and ii) computing ``modeling quality'' typically requires a trial-and-error process whereby only domain-knowledge can be used to discern a good clustering from a bad one. 
In practice, such a scenario illustrates a typical ``development'' workflow which requires: a \emph{batch processing phase} (e.g., running CVB), an \emph{interactive query phase on temporary data} (i.e., on data interesting in relatively short periods of time), and several iterations of both phases until algorithms are properly tuned and final results meet users' expectations.

\db explicitly tackles such ``development'' workflows. Unlike current approaches, which generally require a long and costly data loading phase that considerably increases the data-to-insight time, \db allows querying temporary data in-situ, and exposes a standard SQL interface to the user. This simplifies query analysis and reveals the main advantage of \db in this use case, that is the removal of the \emph{temporary data} loading phase, which today represents one of the main operational bottlenecks in data analysis. 
Indeed, the traditional data loading phase makes sense when the workload (i.e., data and queries) is stable in the long term. However, since data loading may include creating indexes, serialization and parsing overheads,
it is reasonable to question its validity when working with temporary data, as in our machine learning use case.

The key design idea behind \db is that of shifting the part of the burden of a traditional load operation to the batch processing phase of a ``development'' workflow. While batch data processing takes place, \db piggybacks the creation of distributed positional maps and vertical indexes (see Section~\ref{sec:architecture} for details) to improve the performance of interactive user queries on the temporary data. Interactive queries operate directly on temporary data files produced by the batch processing phase, which are stored on a distributed file system such as HDFS \cite{HDFS}.

\subsection{Data exploration}
\label{sec:de}

In this section, we discuss another prominent use case -- which we also evaluate in Section~\ref{sec:eval} -- and which is another important motivation for our work. Here we consider a user 
involved in a preliminary, yet often fundamental and time-consuming, \emph{data exploration} task. Typically, the user collects data from different sources (e.g., an operational system, a public API) and stores it on a distributed file system such as HDFS for subsequent processing. However, before any useful processing can happen, data needs to be ``cleaned'' and studied in detail.

\begin{figure}[t!]
	\center
  \includegraphics[width=1.0\columnwidth]{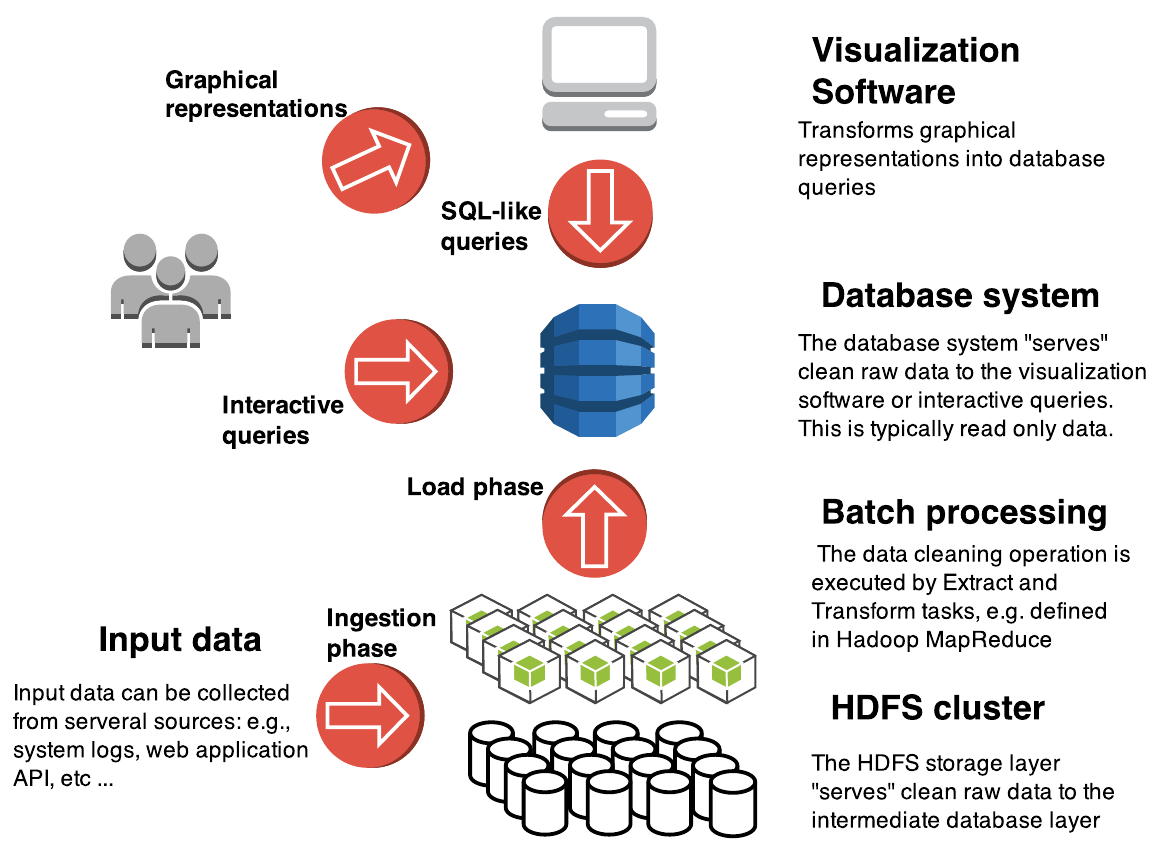}
  \caption{A typical data exploration architecture.}
  \label{fig:data_exploration}
	\vspace{-0.5em}
\end{figure}

Data exploration generally requires visualization tools, that assist users in their preliminary investigation by presenting the salient features of raw data. Current state-of-the-art architectures for data exploration can be summarized as in Figure~\ref{fig:data_exploration}. A \emph{batch processing phase} ingests ``dirty'' data to produce temporary data; such data is then loaded into a database system that supports an \emph{interactive query phase}, whereby a visualization software (e.g., Tableau Software) translates user-defined graphical representations into a series of queries that the database system executes. Such queries typically ``reduce'' data into aggregates, by filtering, selecting subsets satisfying predicates and by taking representative samples (e.g., by focusing on top-$k$ elements).

In the scenario depicted above, \db 
reduces the data-to-insight time, by allowing visualization software or users to directly interact with the raw representation of temporary data, without paying the cost of the load phase that traditional database systems require, nor data format transformation overheads. In addition, the metadata that \db generates by piggybacking on the batch processing phase (while data is ``cleaned''), substantially improves query performance, making it a sensible approach for applications where interactivity matters.

%% file: 03-dinodb.tex
\section{\db architecture and implementation}
\label{sec:architecture}

In this section, we present the architecture design of \db in detail.
\db is designed to provide a seamless integration of batch processing systems such as Hadoop MapReduce and Spark, with a distributed solution for in-situ data analytics on large volumes of temporary, raw data files. 
First, we explain how \db extends the ubiquitous Hadoop I/O API using \plugin, a mechanism that generates a wide range of auxiliary metadata structures to speed-up the interactive data analysis using the \db query engine. 
Then, we describe the \db query engine, which leverages the metadata generated in the batch processing phase to achieve interactive-speed query performance.

In the remaining of this section we assume that both the raw and the \emph{temporary} data ingested and produced by the batch processing phase, and used in the query serving phase are in a structured textual data format (e.g., comma-separated value files).

\subsection{High-level design}
\label{sec:hld}

The batch processing phase (e.g., in the machine learning and data exploration use cases outlined previously) typically involves the execution of (sophisticated) analysis algorithms. This phase might include one or more batch processing jobs, whereby output data is written to HDFS.

The key idea behind \db is to leverage batch processing as a preparation phase for future interactive queries. Namely, \db enriches the Hadoop I/O API with \plugin. Such mechanism piggybacks the generation of metadata by \emph{pipelining} the output tuples produced by the batch engine into a series of specialized \emph{decorators} that store auxiliary metadata along with the original output tuples. We further detail \plugin and metadata generation in Section~\ref{sec:batch}.

In addition to the metadata generation, \db capitalizes on data pre-processing by keeping output data in-memory. To be more specific, we configure Hadoop to store output data and metadata in RAM, using the \texttt{ramfs} file system as an additional mount point for HDFS.\footnote{This technique has been independently considered for inclusion in a recent patch to HDFS \cite{hortonworks} and in a recent in-memory HDFS alternative called Tachyon 
\cite{Tachyon}. Finally it is now added in the latest version of Apache Hadoop \cite{ArchivalStorage}} Our \db prototype supports both \texttt{ramfs} and disk mount points for HDFS, a design choice that allows supporting queries on data that cannot fit in RAM.

Both the output data and metadata are consumed by the \db interactive query engine. As we detail in Section~\ref{sec:mpp}, the \db interactive query engine is a massively parallel processing engine that orchestrates several \db nodes. Each \db node is an optimized instance of PostgresRaw~\cite{Alagiannis:2012:NEQ:2213836.2213864}, a variant of PostgreSQL tailored to querying temporary data files produced in the batch processing phase. 
To ensure high performance and low query execution times, we co-locate \db nodes and HDFS DataNodes, where the two share data through HDFS, and in particular through its in-memory, \texttt{ramfs} mount.

\begin{figure}
	\centering
	\subfigure[\db positional maps]{\includegraphics[width=0.65\linewidth]{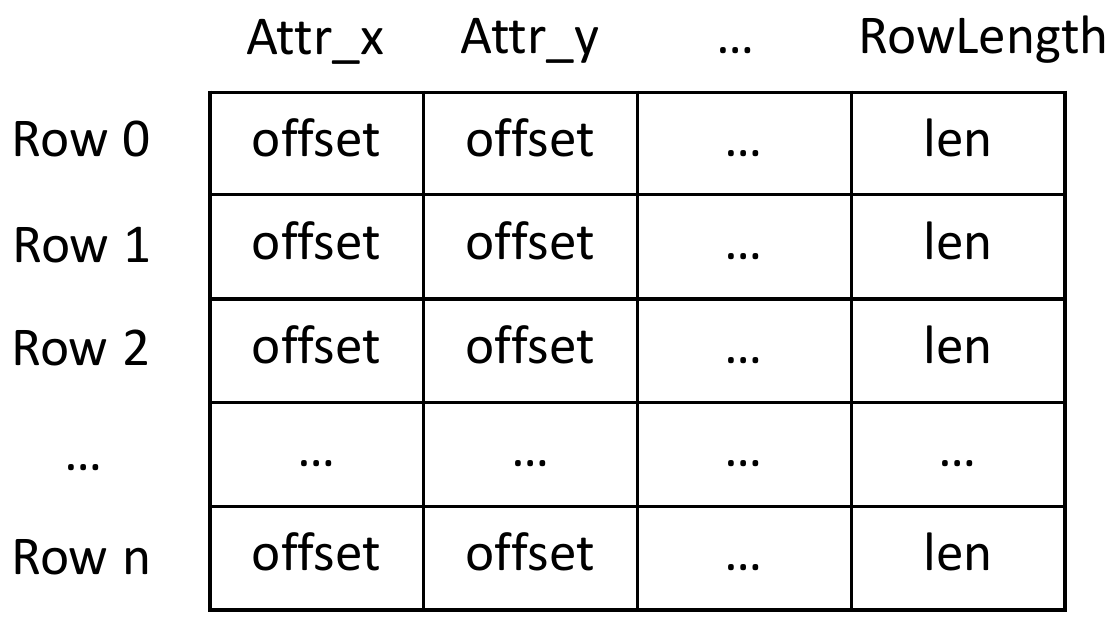}\label{fig:pm}}
	\subfigure[\db vertical indexes]{\includegraphics[width=0.65\linewidth]{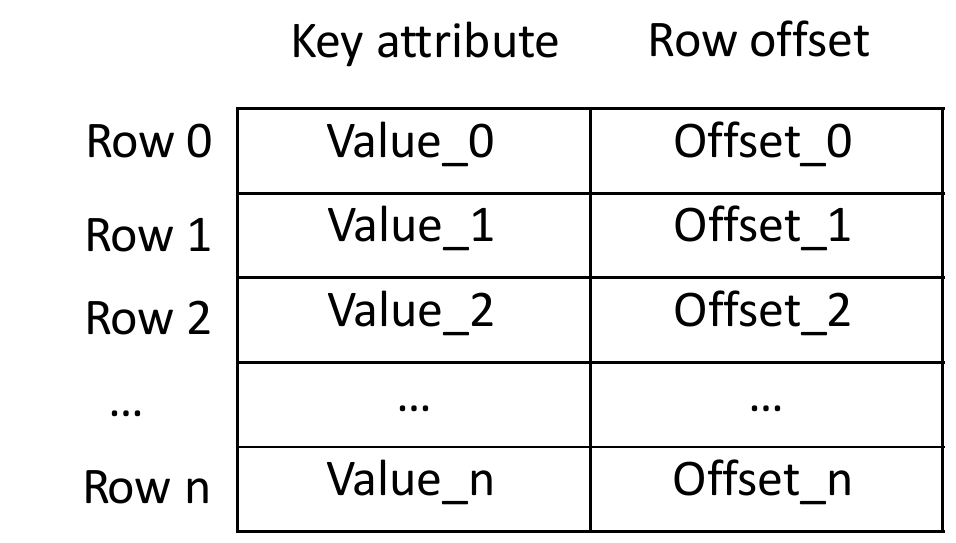}\label{fig:verticalindex}}
	\caption{Metadata generated by \db I/O decorators.}
	\label{fig:dinodb}
	\vspace{-0.5em}
\end{figure}

\subsection{\plugin}
\label{sec:batch}
\db piggybacks the generation of auxiliary metadata on the batch processing phase using \plugin. Next, we introduce metadata currently supported in our prototype, and outline the implementation of our decorators to generate \emph{positional maps}~\cite{Alagiannis:2012:NEQ:2213836.2213864}, \emph{vertical indexes} and \emph{statistics}.

\textbf{Positional maps.} Positional maps are data structures that 
\db uses to optimize in-situ querying. Contrary to a traditional database index, a positional map indexes the structure of the file and not the actual data. A positional map (shown in Figure~\ref{fig:pm}) contains relative positions of attributes in a data record, as well as the length of each row in the data file. During query processing, the information contained in the positional map can be used to jump to the exact position (or as close as possible) of an attribute, significantly reducing the \emph{cost of tokenizing and parsing} when a data record is accessed.

To keep the size of the generated positional map relatively small to the size of a data file, 
the \emph{positional map decorator} implements \emph{uniform sampling}, to store positions only for a subset of the attributes in a file. 
The \emph{positional map decorator} implements \emph{sampling}, to store positions only for a subset of the attributes in a file. The user can either provide a sampling rate so that the positional map decorator will perform uniform sampling, or directly indicate which attributes are sampled.
An approximate positional map can still provide tangible benefits: indeed, if the requested attribute is not part of the sampled positional map, a nearby attribute position is used to navigate quickly to the requested attribute without significant overhead. In Section~\ref{sec:samplepm}, we show the effect of different sampling rates in the query execution performance.

\textbf{Vertical indexes.} 
The positional map can reduce the CPU processing cost associated with parsing and tokenizing data;  
to provide the performance benefit of an index-based access plan, \db uses a \emph{vertical index decorator} that accommodates one or more \emph{key attributes} for which vertical indexes are created at the end of the batch processing phase. Such vertical indexes can be used to quickly search and retrieve data without having to perform a full scan on the temporary output file. Figure~\ref{fig:verticalindex} shows the in-memory data structure of a vertical index. An entry in the vertical index has two fields for each record of the output data: the key attribute value and the record row offset value. 
As such, every key attribute value is associated with a particular row offset in the data file, which \db nodes use to quickly access a specific row of a file.
As decorators generate metadata in a single pass, the key attribute values are not required to be unique or sorted. Each time when the vertical index decorator receives a tuple, it generates the index entry for this tuple which is output to a vertical index file.

\textbf{Statistics.} 
Modern database systems rely on statistics to choose efficient query execution plans. Query optimization requires knowledge about the nature of processed data 
that helps ordering operators such as joins and selections; however, such statistics are available only after loading the data or after a pre-processing phase. 
Currently \plugin can compute the number of records and the number of distinct values for specific attributes from the batch processing phase as the statistics of the output data.
To achieve this, our \emph{statistics decorator} uses the near-optimal probabilistic counting algorithm HyperLogLog~\cite{flajolet2008hyperloglog}. Statistics on attribute cardinality are used by \db to improve the quality of the query plans for complex queries, e.g., involving join operations.
Other kinds of statistics (e.g, skew in the distribution of values per attribute) can be easily supported by \plugin as long as there exist a one-pass algorithm to generate them.

\begin{figure}
	\centerline{\includegraphics 
	[scale=0.53]
	{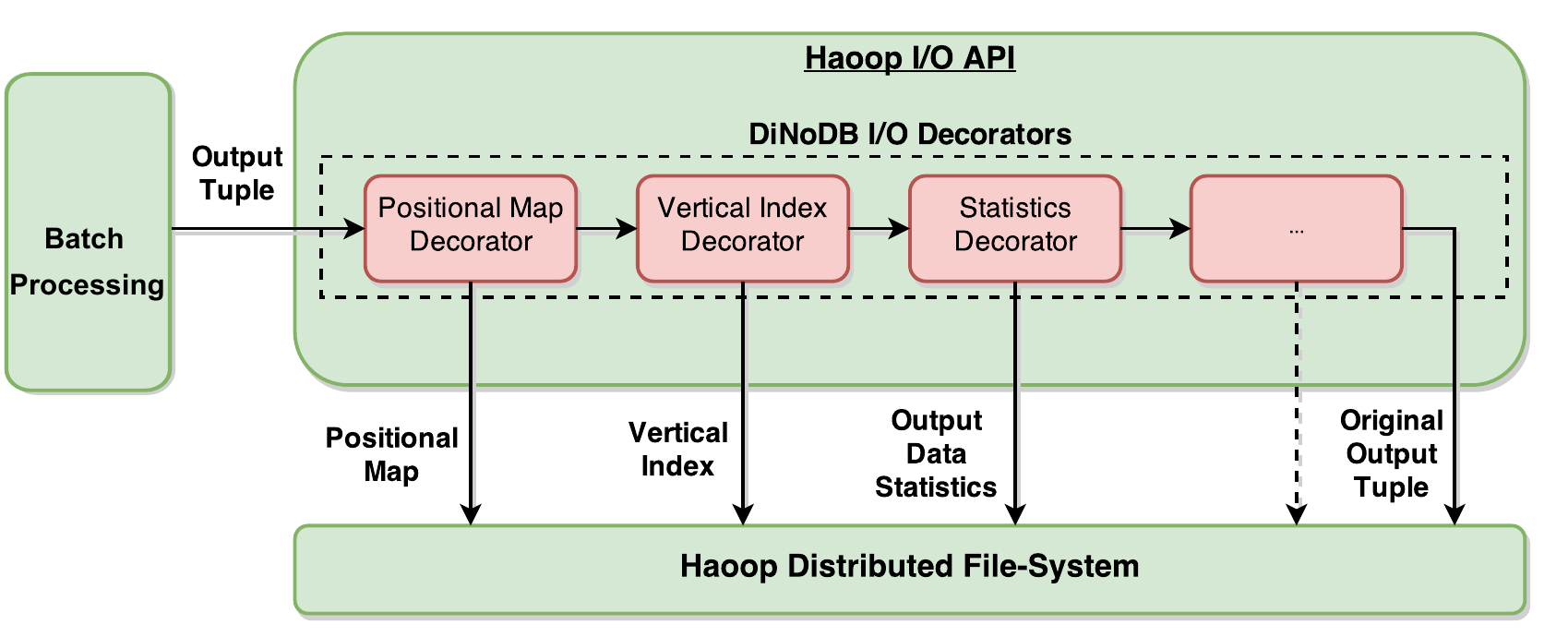}}
	\caption{\db I/O decorator overview.}
	\label{fig:diplugin}
\end{figure}

\begin{algorithm}
\caption{Positional map generation.} \label{alg:generate-pm}
\begin{algorithmic}[1]
\Procedure{initialize}{}
  \State open $pmstream$
\EndProcedure
\Procedure{process}{tuple}
  \State tuple consists of key $attr_0$ and value $[attr_1, attr_2, \cdots, attr_n]$
  \State initialize $tuplestring$, as an empty string
\For{$attr \in [attr_0, attr_1, attr_2, \cdots, attr_n]$}
\If{$attr$ is sampled} 
\State $offset$ $\leftarrow$ getLength($tuplestring$) \; 
\State $pmstream$.write($offset$) \;
\EndIf
\State $tuplestring \leftarrow tuplestring + ',' + attr$ \;
\EndFor
\State $len$ $\leftarrow$ getLength($tuplestring$) \; 
\State $pmstream$.write($len$) \;
\State tuple.setString($tuplestring$)
\State pass tuple to next decorator
\EndProcedure
\Procedure{close}{}
  \State close $pmstream$
\EndProcedure
\end{algorithmic}
\end{algorithm}

\textbf{Implementation details.} 
\plugin are designed to be a non-intrusive mechanism, that seamlessly integrates systems supporting the classical Hadoop I/O API, such as Hadoop MapReduce and Apache Spark.

\plugin operate at the end of the batch processing phase for each final task that produces output tuples, as shown in Figure~\ref{fig:diplugin}. Instead of writing output tuples to HDFS directly, using the standard Hadoop I/O API, the tasks use \plugin, which build a metadata generation pipeline, where each decorator iterates over streams of output tuples and compute the various kinds of metadata described above.
For example, to generate positional maps, the positional map decorator executes Algorithm~\ref{alg:generate-pm}:
The positional map decorator is initialized by opening a positional map file  stream $pmstream$ (line 1-3); 
the decorator continuously receives tuples from which lists of attributes can be extracted (line 5);
for each tuple, the decorator constructs string $tuplestring$ by iterating through all attributes (line 7-13); during string construction, the offsets of sampled attributes are written to $pmstream$ (line 8-11);
when $tuplestring$ is fully formed, its length is also written to $pmstream$ (line 14-15); then the original tuple and its $tuplestring$ is passed to the next decorator (line 16-17); 
when all tuples are processed, $pmstream$ is closed so that the positional map file is finalized (line 19-21).

To use \plugin, Hadoop users need to replace the vanilla Hadoop \texttt{OutputFormat} class by a new module called \texttt{DiNoDBOutputFormat}. Our prototype currently supports the \texttt{TextOutputFormat} sub-class, which allows \db to operate on textual data formats. Specifically, the \texttt{DiNoDBTextOutputFormat} module implements a new \texttt{DiNoDBArrayWritable} class which is used to generate both the 
output data and its associated metadata. 
If users work with Spark, before saving their result \texttt{RDD} to HDFS (by method \texttt{saveAsTextFile}) they need to first cast that result \texttt{RDD} to a \texttt{DiNoDBRDD}, which internally uses \texttt{DiNoDBOutputFormat} as \texttt{OutputFormat} class.

\db I/O decorators are configured by passing a configuration file to each batch processing job in Hadoop or 
by setting parameters of
\texttt{DiNoDBRDD} in Spark. Users specify which metadata to generate and indicate parameters, such as the sampling rate to use for the generation of positional maps and the key attributes for the generation of vertical indexes.

\textbf{Discussion.}
Although currently DiNoDB focuses on textual
data format, the same idea of generating metadata could also be applied to
other data formats, like binary files. Depending on different input data format, generated metadata may be different. For example, if the data is in FITS ~\cite{fits} data format,
positional map is not needed anymore because each tuple and attribute is usually located in a well-known location. However, vertical indexes and statistics would 
still help.

\subsection{The \db interactive query engine}
\label{sec:mpp}

\begin{figure}
	\centerline{\includegraphics[scale=0.65]{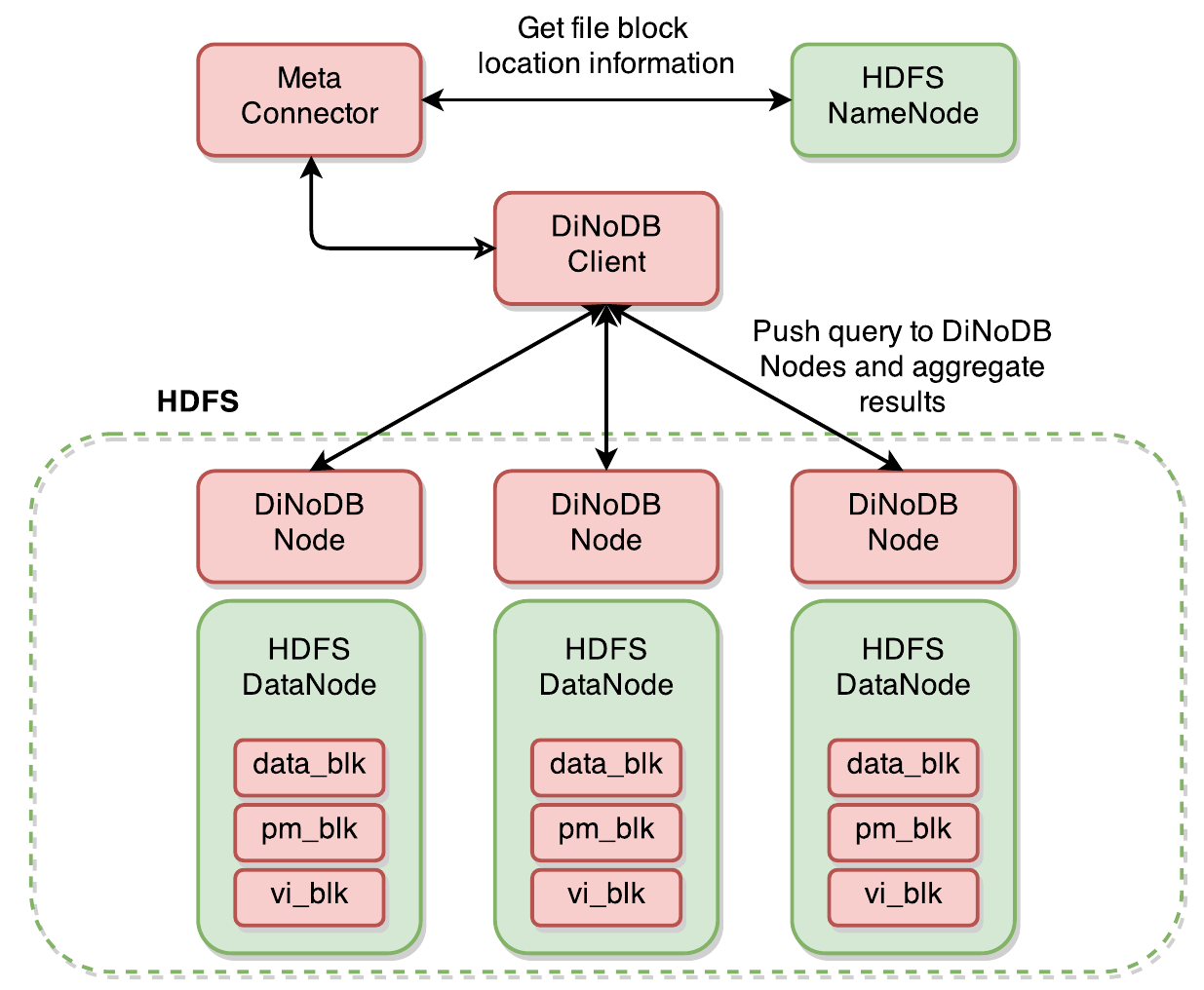}}
	\caption{Architecture of the \db interactive query engine.}
	\label{fig:dinodbarc}
\end{figure}

At a high level (see Figure~\ref{fig:dinodbarc}), the \db interactive query engine consists of a set of \db nodes, orchestrated using a  massively parallel processing (MPP) framework. In our prototype implementation, we use the Stado MPP framework \cite{stado}, which nicely integrates PostgreSQL-based database engines.
\db ensures data locality by co-locating \db nodes with HDFS DataNodes.

In the following sections, we first describe the \db client (Section~\ref{sec:clients}) and the \db nodes (Section~\ref{sec:dinodbnode}). Then, we describe how \db achieves fault-tolerance (Section~\ref{sec:FT}).

\subsubsection{\db clients}
\label{sec:clients}

A \db client serves as entry point for \db interactive queries. It provides a standard shell command interface, hiding the network layout and the distributed system architecture from the users. As such, applications can use \db just like a traditional DBMS.

\db clients accept application requests (queries), and communicate with \db nodes. When a \db client receives a query, it fetches the metadata for the ``tables'' (output files of the batch phase) indicated in the query, using the \emph{MetaConnector} module. The MetaConnector (see Figure~\ref{fig:dinodbarc}) operates as a proxy between \db and the HDFS NameNode, and is responsible for retrieving HDFS metadata information like partitions and block locations of raw data files.
Using HDFS metadata, the MetaConnector guides the \db clients to query the \db nodes that hold raw data files relevant to the user queries. Additionally, the MetaConnector remotely configures \db nodes so that they can build the mapping between ``tables'' and the related HDFS blocks, including all data file blocks, positional map blocks and vertical index blocks. In summary, the anatomy of a query execution is as follows: i) using the MetaConnector, a \db client learns the location of every raw file block and pushes the query to the respective \db nodes; ii) the \db nodes process the query in parallel; and finally, iii) the \db client aggregates the result.

Note that, since the \db nodes are co-located with the HDFS DataNodes, \db inherits fault-tolerance from HDFS replication. If a \db client detects a failure of a \db node, or upon the expiration of a timeout on \db node's responses, the \db client will issue the same query to another \db node holding a replica of the target HDFS blocks. We discuss \db fault-tolerance in more details in Section~\ref{sec:FT}.

\subsubsection{\db nodes}
\label{sec:dinodbnode}

The \db nodes are based on PostgresRaw~\cite{Alagiannis:2012:NEQ:2213836.2213864}, a query engine optimized for in-situ querying. In the following, we first briefly recall how PostgresRaw differs from native PostgreSQL and then explain all the details behind \db nodes, and differences with respect to PostgresRaw.

\textbf{PostgresRaw.} PostgresRaw is a centralized instantiation of the NoDB paradigm~\cite{Alagiannis:2012:NEQ:2213836.2213864}, and is a variant of PostgreSQL that avoids the data loading phase and executes queries directly on data files. 
PostgresRaw adopts in-situ querying instead of loading and preparing the data for queries. To accelerate query execution, PostgresRaw tokenizes only necessary attributes and parses only qualified tuples. Moreover, PostgresRaw \emph{incrementally} builds a positional map, which contains relative positions of attributes in a line, and updates it during the query execution: as such, the more queries are executed, the more complete and useful (for performance) the positional map will be.

\textbf{From PostgresRaw to \db nodes.} \db nodes instantiate customized \nodb databases which execute user queries, 
and are co-located with HDFS DataNodes. In the vanilla \nodb \cite{Alagiannis:2012:NEQ:2213836.2213864} implementation, a ``table'' maps to a single data file. Since the HDFS files are instead split into multiple blocks, \db nodes use a new file reader mechanism that 
can access data on HDFS and maps a ``table'' to a list of data file blocks. In addition, the vanilla \nodb implementation is a multiple-process server, which forks a new process for each new client session, with individual metadata and data cache per process. Instead, \db nodes place metadata and data in \emph{shared memory}, such that user queries -- which are sent through the \db client -- can benefit from them across multiple sessions.

\db nodes can take advantage of the fact that data is naturally partitioned into HDFS blocks to leverage modern multi-core processors. 
Hence, data and the associated metadata can be easily accessed by multiple instances of \nodb, to allow node level parallelism.
\db users can selectively indicate whether raw data files are placed on disk or in memory. Hence, \db nodes can seamlessly benefit from a memory-backed file system to dramatically improve query execution times.

\textbf{Exploiting metadata.} 
\db nodes leverage positional map files generated in the pre-processing phase when executing queries, instead of only populating them incrementally as in \nodb. In the case of the approximate positional maps, \db nodes use the sampled attributes as anchor points, to retrieve nearby attributes 
within the same row, required to satisfy a query.
When vertical indexes are available, \db nodes use them to speed up queries with low selectivity, by employing an index-based access plan as a replacement for a full sequential scan. 
Both the positional map and the vertical index files are loaded by a \db node when the first query requires them. As our performance evaluation shows (see Section~\ref{sec:eval}), the metadata loading time is almost negligible, when compared to the execution time of the query.

\textbf{Data update.} 
If new data is injected to an existing table without using \plugin (e.g., manually upload), there is no associated metadata pre-generated. 
In this case, \db nodes can still exploit the available metadata and do not require new metadata for the newly added data to process queries.
Partially-available metadata 
can still accelerate query execution.
The missing metadata can be generated incrementally after a few queries
and kept in \db nodes' memory.
HDFS is an immutable filesystem, which means that both data and metadata can not be modified after being written. If part of data is deleted, the associated metadata needs to be deleted as well.

\subsubsection{Fault tolerance}
\label{sec:FT}

In a nutshell, the key idea behind \db fault tolerance is to exploit HDFS n-way replication, in which every HDFS block on a given node is replicated to n-1 other nodes. As \db nodes co-locate with HDFS DataNodes, user queries can be directed to multiple nodes: in case one node is not available, \db clients automatically forward queries to other nodes with replicas of the data. By virtue of pro-active request redirection, \db can address the issues related to latency-tail tolerance \cite{Dean:2013:TS:2408776.2408794}. 

However, the default HDFS replication mechanism is not suited for \db: indeed, HDFS does not guarantee data and metadata generated by \plugin to be replicated on the same nodes. For example, some blocks assigned to DataNode $D_1$ may be replicated across DataNodes $D_2$ and $D_3$, whereas other blocks assigned to $D_1$ might be replicated differently, e.g., on DataNodes $D_4$ and $D_5$.

We thus proceed with the design of a new replication mechanism for HDFS that is tailored to \db, which achieves two objectives: (i) it co-locates data blocks with the corresponding metadata blocks and (ii) it allows selecting different ``storage levels'' for replicas, to save on cluster resources. 
To address data/metadata co-location, our \db prototype implements a \emph{per-node} n-way replication. Every block assigned to a given DataNode $D_i$ is systematically replicated across the same DataNodes $D_j$ and $D_k$. We are aware that, on the long run, such simple approach may lead the HDFS subsystem to be poorly balanced. An alternative approach that we are currently considering is to create a new Hadoop Output Format that, similarly to Apache Parquet \cite{parquet}, supports ``containers'' that include data and metadata.
Finally, \db supports different storage levels across replicas: as such, a ``primary'' replica can be tagged to be stored in the HDFS \texttt{ramfs} mount point, while ``secondary'' replicas are instead stored in an HDFS disk-based mount point \cite{krish2014hats}.

%% file: 04-experiments.tex
\section{Experimental evaluation}
\label{sec:eval}

\begin{table*}[t]
\renewcommand{\arraystretch}{1.3}
\centering
\caption{Comparison of Systems}
 \begin{tabular}{ | c | c | c | c | c | c | c | c |}
   \hline
\textbf{Systems} & \textbf{DiNoDB} & \textbf{ImpalaT} & \textbf{ImpalaP} & \textbf{SparkSQL} & \textbf{SparkSQLc} & \textbf{Hive on Tez} & \textbf{Stado} \\  
   \hline
   Loading requirement & no & no & yes& no & yes & no & yes \\ \hline
   Indexing support & \shortstack{yes, by pre-\\generated index} & no & column-oriented & no & no & no & yes \\
   \hline
 \end{tabular}
 \label{table:comparison}
 \end{table*}

We now present a detailed experimental evaluation of \db using both real and synthetic datasets. 
We compare \db against state-of-the-art data analytics systems, 
including Impala, SparkSQL, Hive on Tez and Stado. Stado, an MPP system based on PostgreSQL, needs to load 
the data before executing queries, while Hive on Tez executes queries directly on data files stored in HDFS. 
Impala and SparkSQL can execute queries either directly on data files or after 
creating a copy of data in a new file format such as Parquet.
Additionally, SparkSQL provides functionality to cache data files in memory
as Resilient Distributed Dataset (RDD) \cite{Zaharia:2012:RDD:2228298.2228301}.

Our goal here is to perform a comparative analysis across all systems 
showing the \emph{aggregate query execution time}, which accounts for the total time to execute a certain number 
of queries on temporary output data of the batch processing phase.
Additionally, we highlight the merits and applicability of metadata generated by the \plugin.

\subsection{Experimental Setup}
\label{sec:setup}

All the experiments are conducted in a cluster of 
9 machines, with 4 cores, 16 GB RAM and 1 Gbps network interface each. 
The underlying distributed file system is HDFS. 
Eight machines are configured as worker nodes (DataNode), 
while the remaining one is acting as a coordinator (NameNode) in HDFS. 
To avoid disk bottlenecks, 
all the datasets are stored in HDFS with a \texttt{ramfs} mount point on each DataNode.

We compare \db with Impala (version 1.4), SparkSQL (version 1.1.0), 
Hive (version 0.13.1) on Tez (version 0.4.1) and PostgreSQL-based Stado. 
For \db, we assign 3 out of 4 cores to a \db node and 
we balance the data of the underlying HDFS file system across the 3 cores. 
The fourth core acts as a server and coordinates the three client-sessions of PostgresRaw.
All the other systems also fully utilize CPU resources, either by multithreading (Impala and SparkSQL) or by multiprocessing (Hive on Tez and PostgreSQL-based Stado).
In our experiments, we evaluate SparkSQL in two ways: 
i) by caching the entire table in memory as RDD 
before query execution (we label this variant of SparkSQL as SparkSQLc); 
and ii) the ``normal'' execution without caching (labeled as SparkSQL). 
Similarly, we evaluate Impala in two ways: 
i) by converting the dataset from text format to Parquet format before query execution (labeled as ImpalaP); 
and ii) the ``normal'' execution without converting (labeled as ImpalaT). 
A brief comparison among these systems can be found in Table~\ref{table:comparison}.

\subsection{Experiments with synthetic data}
\label{sec:synthetic}
In this section, we use a synthetic dataset of 70 GB containing $5 * 10^{7}$ tuples. 
Each tuple has 150 attributes with integers uniformly distributed in the range [0-$10^{9}$). 
The \plugin already produce a 3.5 GB positional map file (with 1/10 sampling rate) 
and a 1.1 GB vertical index file, where we choose the first attribute as the key attribute. 

We use as input a sequence of SELECT PROJECT SQL queries 
that effectively simulate the kind of workload in a data exploration use case, 
which involves mainly filtering and selecting subsets of data.
We discuss the exact queries in more detail in each experiment.

\begin{figure}
\centerline{\includegraphics[width=1\linewidth]{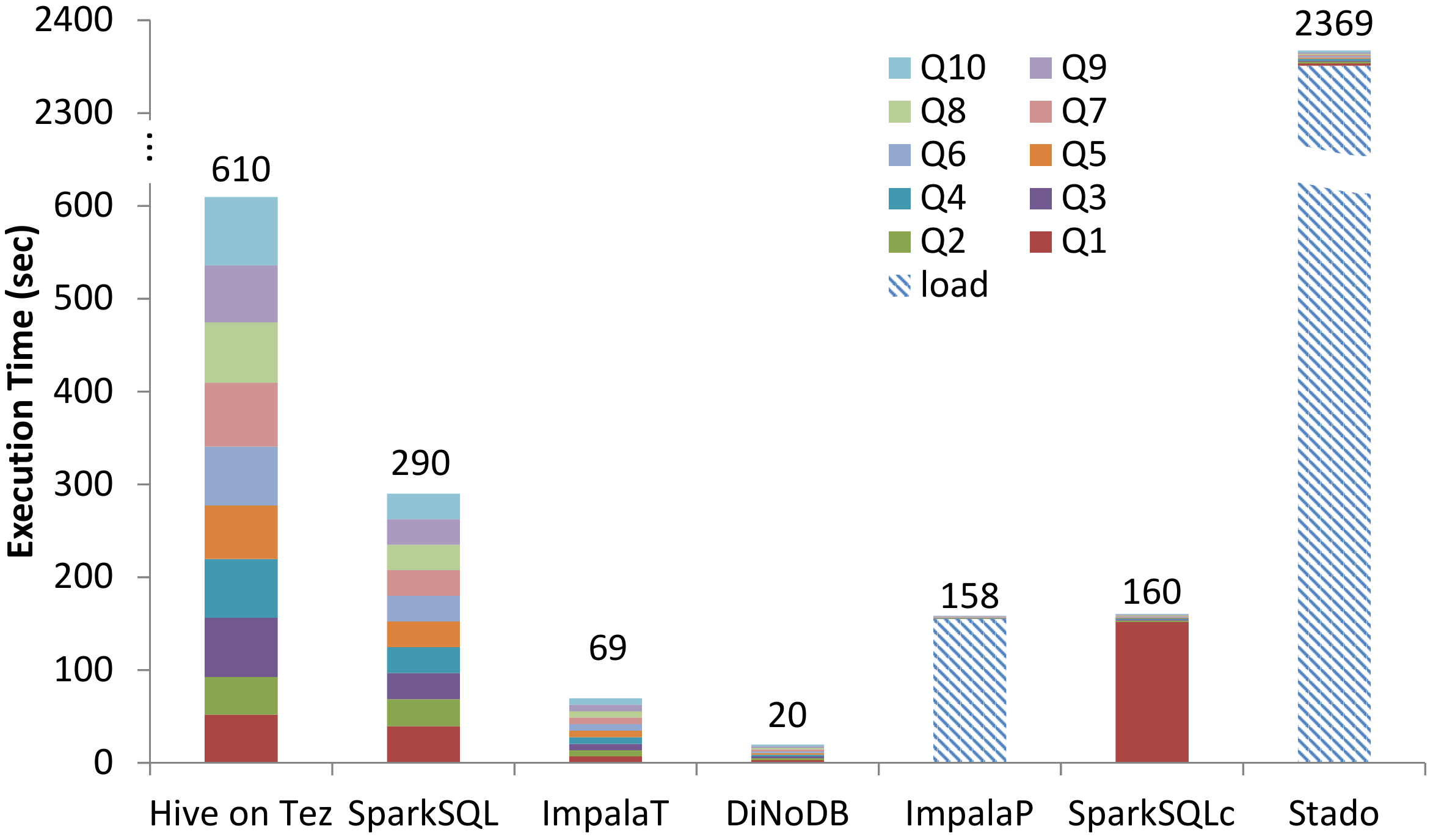}} 
\caption{\db vs. other distributed systems: 
Positional map reduces the cost of accessing data files.}
\label{fig:distributedtest}
\end{figure}

\subsubsection{Random queries (stressing PM)}
\label{sec:randomqueries}
In this experiment, we show the benefit of using positional maps in \db. 
We use as input a sequence of 10 
SELECT PROJECT SQL queries of the following template: 
\texttt{select ax from table where ay < 100000}. Attributes \emph{ax} and \emph{ay} are randomly 
selected and the selectivity is 0.1\textperthousand\xspace.
We examine two categories of systems that evaluate queries i) directly on data files and ii) on loaded data.
Specifically, Hive on Tez, SparkSQL, ImpalaT and \db execute queries directly, without any data loading process.
ImpalaP and Stado require to load the data before querying, while SparkSQLc uses the 
first query to load the data in a memory-backed RDD (lazy loading).

Figure~\ref{fig:distributedtest} plots the query execution time of the 10 queries. For queries over loaded 
data we also report the required loading time. Considering the aggregate query execution time for the 
10 queries, \db is more than three times faster than the second fastest system, ImpalaT.
Additionally, when it comes to individual query times \db consistently outperforms systems executing 
queries on data files. \db achieves that by exploiting the positional map that is generated by 
the \plugin to reduce the CPU cost of accessing data files (parsing and tokenizing).
On the other hand, SparkSQLc, ImpalaP, and Stado achieve shorter query execution times only after spending 
150, 155 and 2352 seconds, respectively, for data loading. 
In Section~\ref{sec:break_even}, we further investigate the trade-off between initially investing time
to prepare the data for querying versus quickly accessing the data on files.

\begin{figure}
\centerline{\includegraphics[width=1\linewidth]{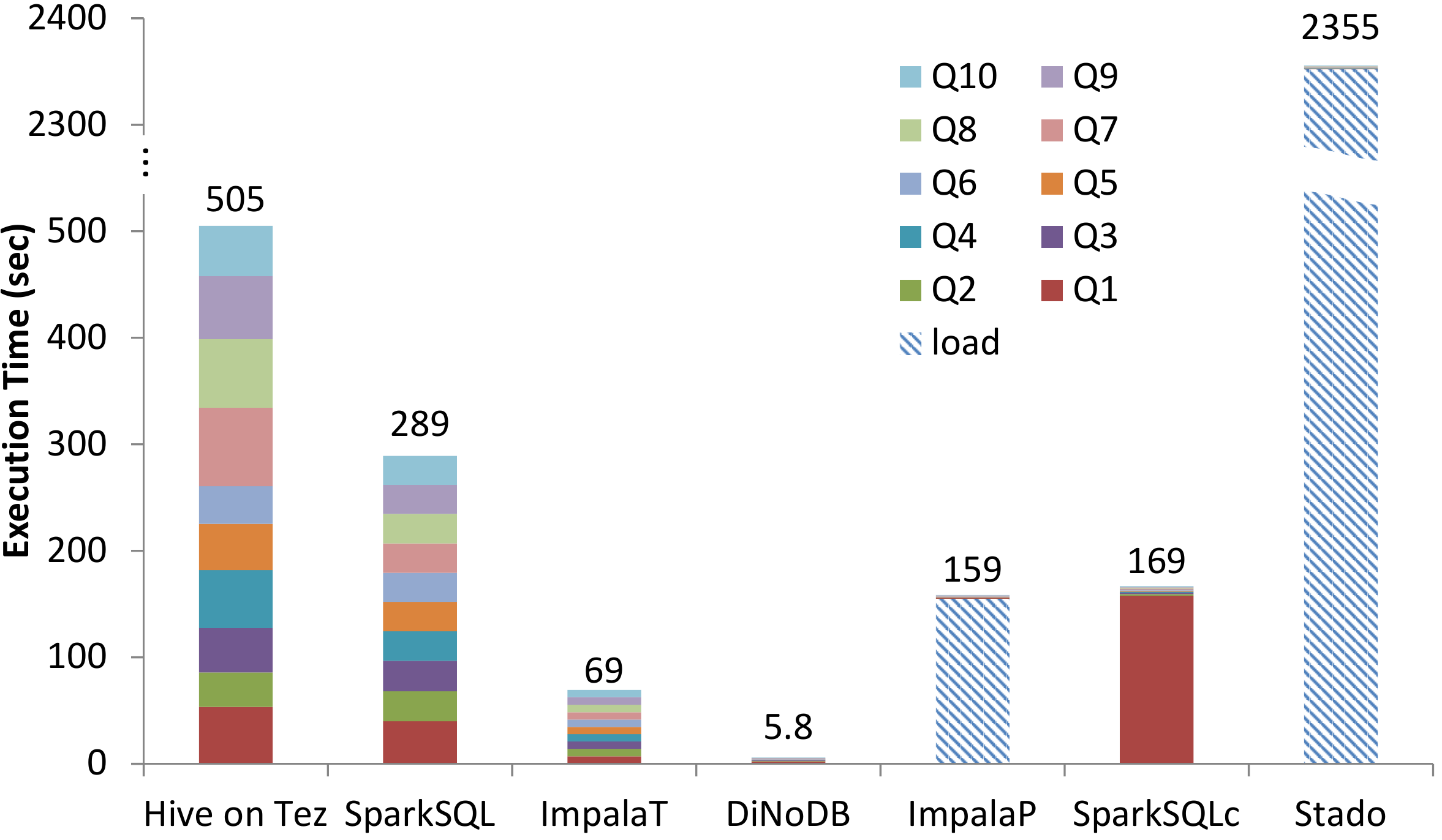}}
\caption{\db vs. other distributed systems: Vertical indexes significantly improve \db's 
performance.}
\label{fig:distributedindextest}
\end{figure}

\subsubsection{Key attribute based queries (stressing VI)}
In this experiment, we demonstrate the impact of exploiting the vertical indexes in \db. 
We use as input a sequence of 10 SELECT PROJECT SQL queries following the 
template: \texttt{select ax from table where akey < 100000}.
The attribute in the WHERE clause is no longer a random attribute, but the attribute which we set 
as the \emph{key attribute} by an appropriate configuration of the \plugin. 
The selectivity is again 0.1\textperthousand\xspace.

Figure~\ref{fig:distributedindextest} shows the query execution time of the query sequence.
\db significantly benefits from exploiting the generated VI file to perform index scan access to the 
data (saving CPU and I/O cost). Overall, the average query cost of \db is less than 1 second. 
On the other hand, Hive on Tez, SparkSQL and ImpalaT that do not have an indexing mechanism, have similar 
performance as in the random query experiment above. 
Since Stado uses PostgreSQL as the local database system in our experiments, 
it supports index scan on data. 
However, 
it needs an additional 9 seconds delay for the index building phase, 
so in total (including the data loading phase), Stado needs about 80 seconds before it is able to execute any queries.

Vertical indexes dramatically accelerate query execution 
in \db if queries hit the key attribute. 
In this experiment, \db's query execution time is competitive compared to Stado, 
ImpalaP and SparkSQLc (besides the first query which is very slow for SparkSQLc). 
However, unlike those systems, \db does not require any data loading phase.
Again, considering the aggregate query execution time for 10 queries, 
\db is more than 10 times faster than the second fastest system we study, ImpalaT.

\begin{figure}
\centerline{\includegraphics[width=\linewidth]{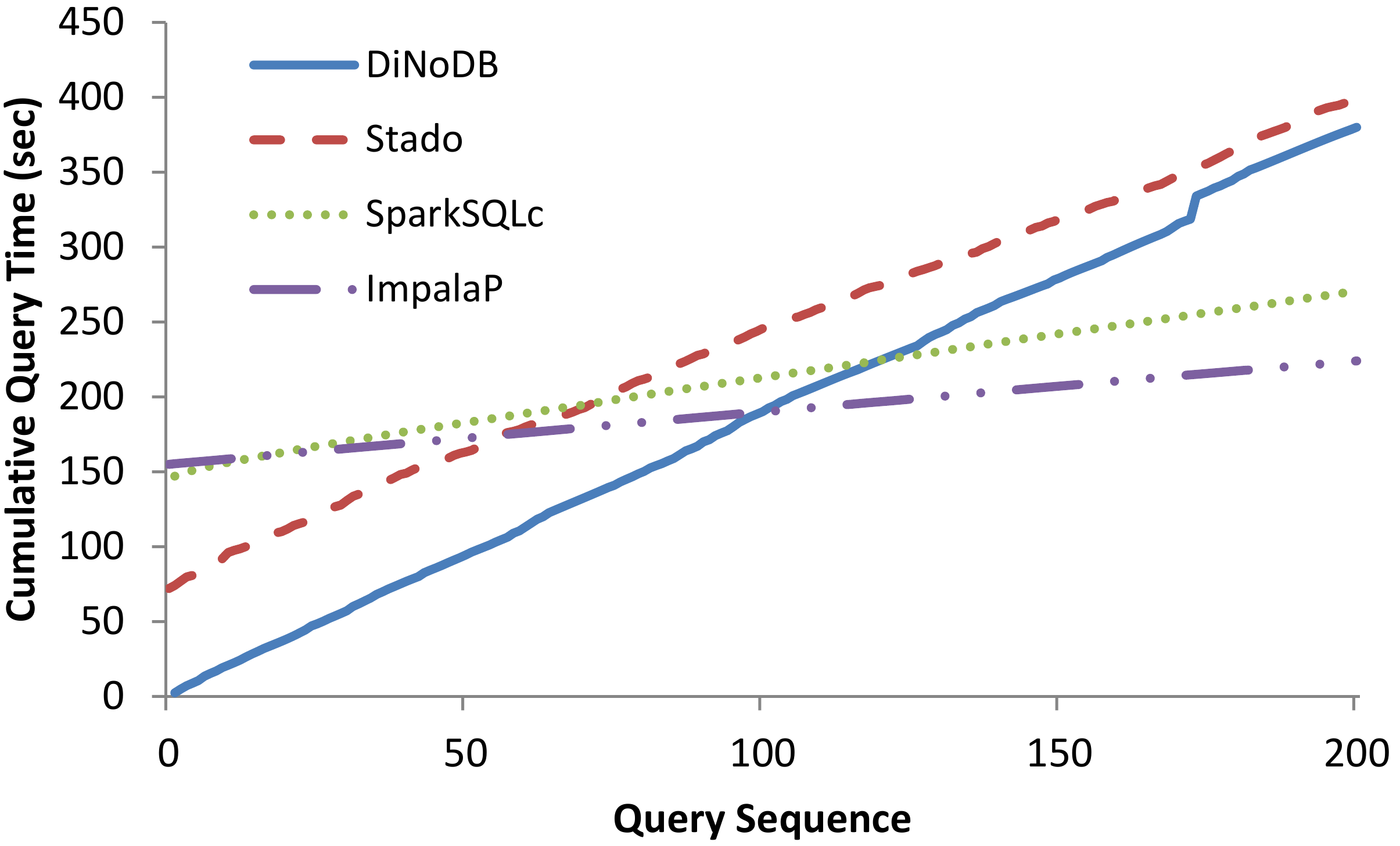}}
\vspace{5.30pt}
\caption{\db is a sensible alternative for data exploration scenarios even for a long sequence of queries.}
\label{fig:queries200}
\end{figure}

\subsubsection{Break-even point}
\label{sec:break_even}

In the previous experiments we show that for a relatively low number of queries, \db outperforms other distributed systems.
In this experiment, we are interested in finding the break-even point, that is the number of queries \db can execute before its performance becomes equal or worse than the alternative systems. To this end, we use the same dataset and the same query patterns as 
in Section~\ref{sec:randomqueries}, for a sequence of 200 queries. We compare \db with Stado, SparkSQLc and ImpalaP. 

The results, shown in Figure~\ref{fig:queries200},
indicate that for this workload, 100 queries represent the break-even point. If users execute less than 100 queries on temporary data, \db outperforms alternative approaches. For more than 100 queries, ImpalaP and SparkSQLc perform better than \db. In this case, the cost of converting the initial dataset to Parquet and RDD, respectively, is amortized.
Clearly, a large number of queries on temporary output data as part of the illustrative exploratory use cases we consider in this work, is unlikely, making \db a sensible choice.

\subsubsection{Impact of data format}
\label{sec:data format}

In the previous experiments we compared different data analytics systems assuming native, uncompressed, data format such as text-based CSV (or JSON). For Impala, in particular, we considered the transformation from text-based to the columnar Parquet format (ImpalaP) as the ``loading'' phase of ImpalaP. Although we expect our assumption of raw, text-based input to be reasonable for \emph{temporary data} of various applications (with such temporary data being the focus of this paper and DiNoDB), one may argue that the comparison to ImpalaP has not been entirely ``fair'' so far --- as a user may write the output of a Hadoop job directly in Parquet format making it ready for ImpalaP and hence avoiding the need for loading/conversion phase.

Here, we compare the raw performance of ImpalaP (with data already in Parquet format) and DiNoDB (with data in text-based CSV format), with ImpalaT (using also text format) as a reference. We run a sequence of SELECT PROJECT queries, varying the number of projected attributes from 1 to 150 (i.e., all attributes, where the query boils down to a 
``\texttt{select * from table where ax < 100000}" 
query), while keeping the  0.1\textperthousand\xspace selectivity and running 50 queries per experiment. 

\begin{figure}
	\centerline{\includegraphics[width=1\linewidth]{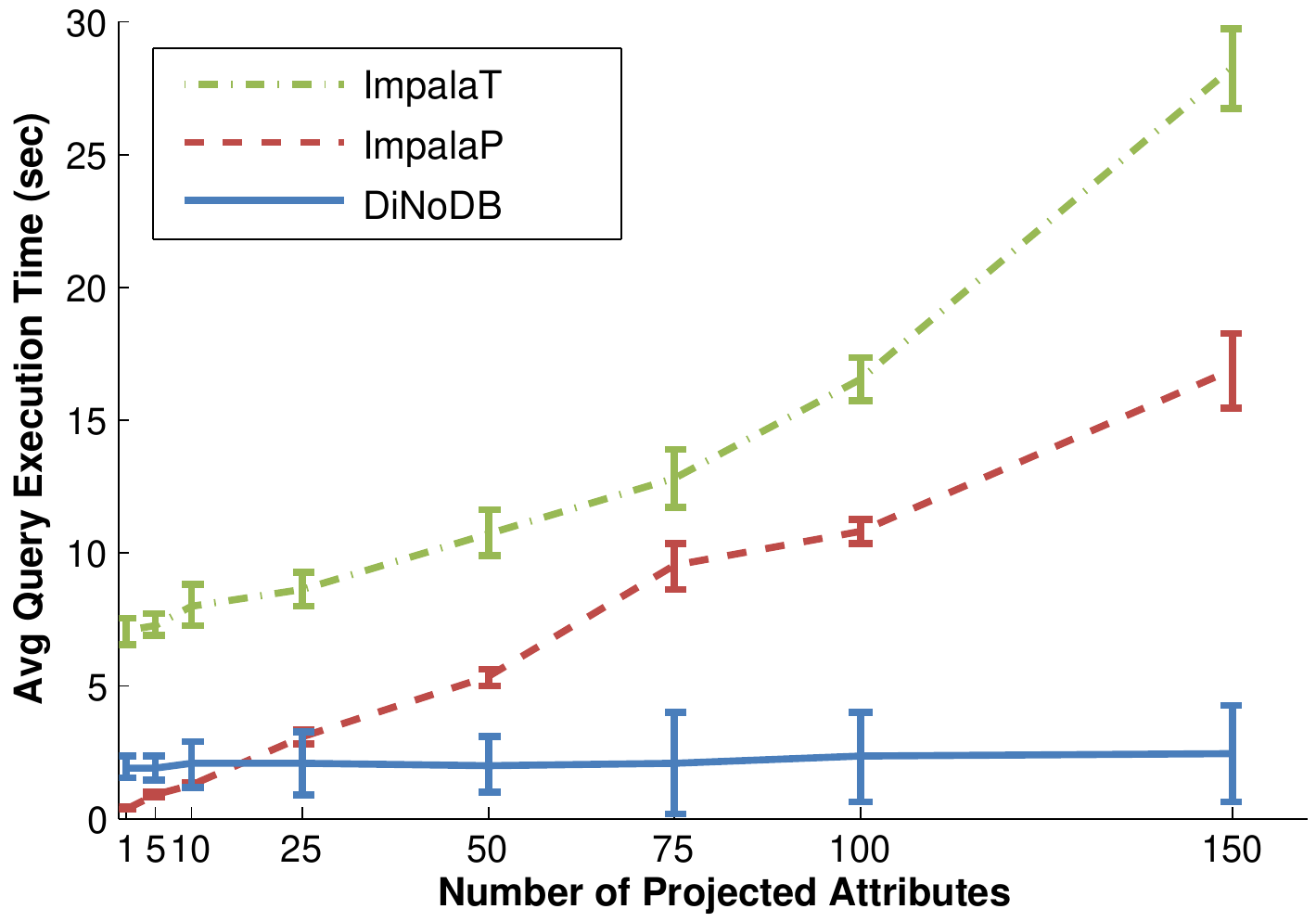}}
	\caption{Latency in function of the number of projected attributes.}
	\label{fig:wearebetterafterall}
\end{figure}

Figure~\ref{fig:wearebetterafterall} shows the average query latency with standard deviation against the number of projected attributes. This experiment shows that ImpalaP and DiNoDB consistently outperform ImpalaT, but that the comparison between ImpalaP and DiNoDB is not conclusive. ImpalaP outperforms DiNoDB for a relatively small number of projected attributes (i.e., less than 30 attributes), whereas DiNoDB has the edge when the number of projected attributes is larger. We can also observe that the performance of DiNoDB is relatively constant with respect to the number of projected attributes, which makes it suitable for a wide range of ad-hoc queries. 
Besides pre-generated metadata, another advantage of DiNoDB comes from the feature inherited from PostgresRaw \cite{Alagiannis:2012:NEQ:2213836.2213864} called \emph{selective parsing}: 
DiNoDB delays the binary transformation of projected attributes until it knows that the selectivity condition is satisfied for the given tuple.
Since the selectivity of queries is very low, DiNoDB significantly reduces the CPU processing costs compared with ImpalaT and ImpalaP.

\subsubsection{Impact of approximate positional maps}
\label{sec:samplepm}

\begin{figure}[!t]
\centerline{\includegraphics[width=\linewidth]{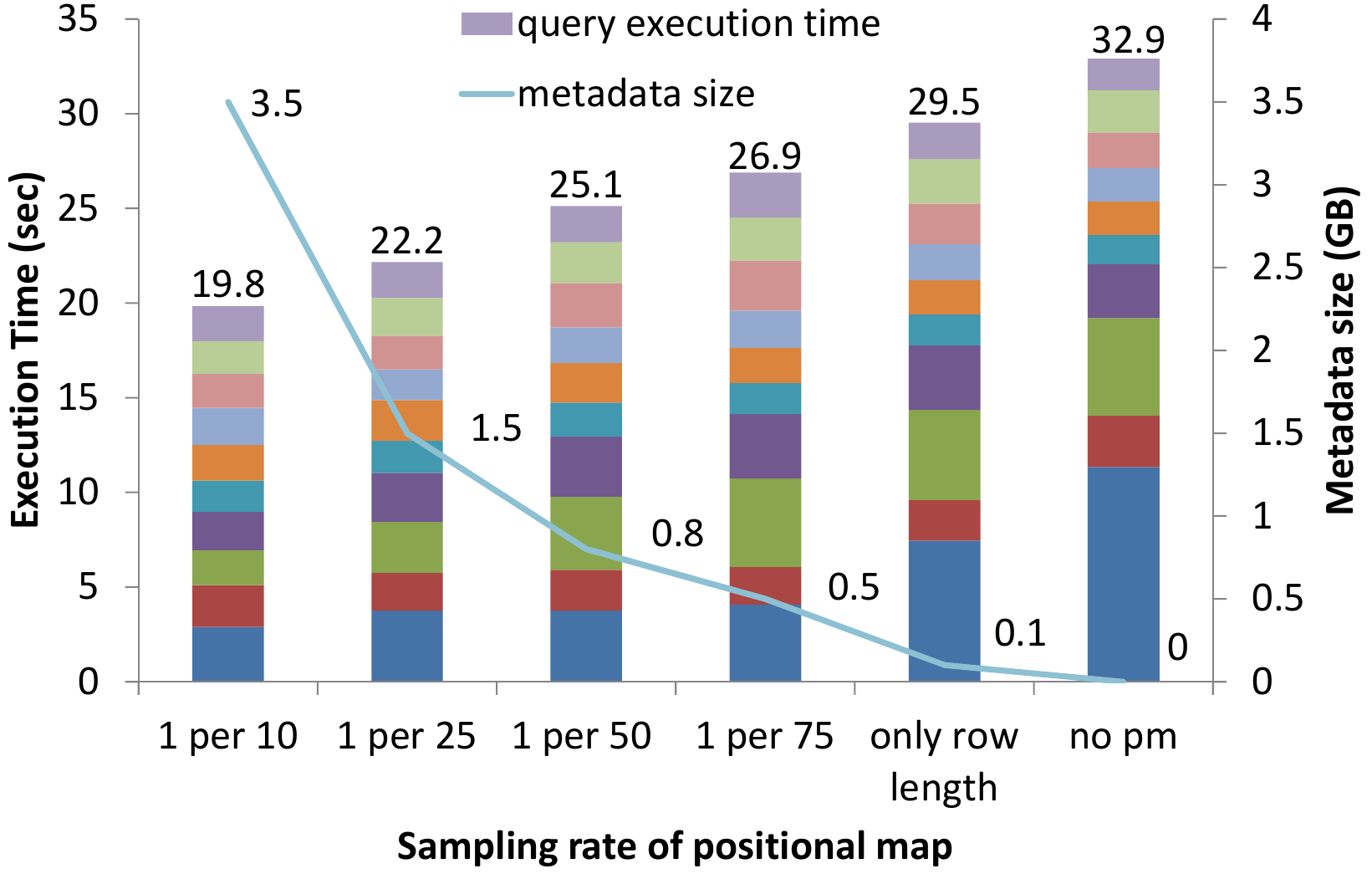}}
\vspace{-3.8pt}
\caption{Different sampling rate of positional map.}
\label{fig:pmsample}
\end{figure}

As we described in Section~\ref{sec:batch}, we use a \emph{uniform sampling} technique to keep PM metadata small. We investigate how different sampling rates of PM influence query execution cost. 
We use the same dataset and query sequence as in Section~\ref{sec:randomqueries}. 
We vary the sampling rate of PM from 1/10, 1/25, 1/50, 1/75, 0 (only containing the length of row) to no PM at all.

The results, shown in Figure~\ref{fig:pmsample}, illustrate the trade-off between query execution time and space constraints.
With more positions kept in the PM, \db achieves lower query execution times, but the metadata file of PM is larger. If a user is more sensitive to query execution time, she can choose a more aggressive sampling rate of PM. Otherwise, if cluster resources are limited, users can choose lower sampling rates, which result in smaller PM files.
Besides pre-generated PM metadata, DiNoDB also incrementally generates PM during query execution. Therefore, as shown in Figure~\ref{fig:pmsample}, the difference in query execution time is more significant in the first few queries when PM is not detailed enough.

\subsubsection{Scalability}
\label{sec:scalability}

In this section, we study the scalability of \db, compared to ImpalaT. Both systems perform a sequential scan operation (i.e., in case \db uses the positional map but not the vertical index).

\begin{figure}
\subfigure[scaling the number of attributes]{\includegraphics[width=\linewidth]{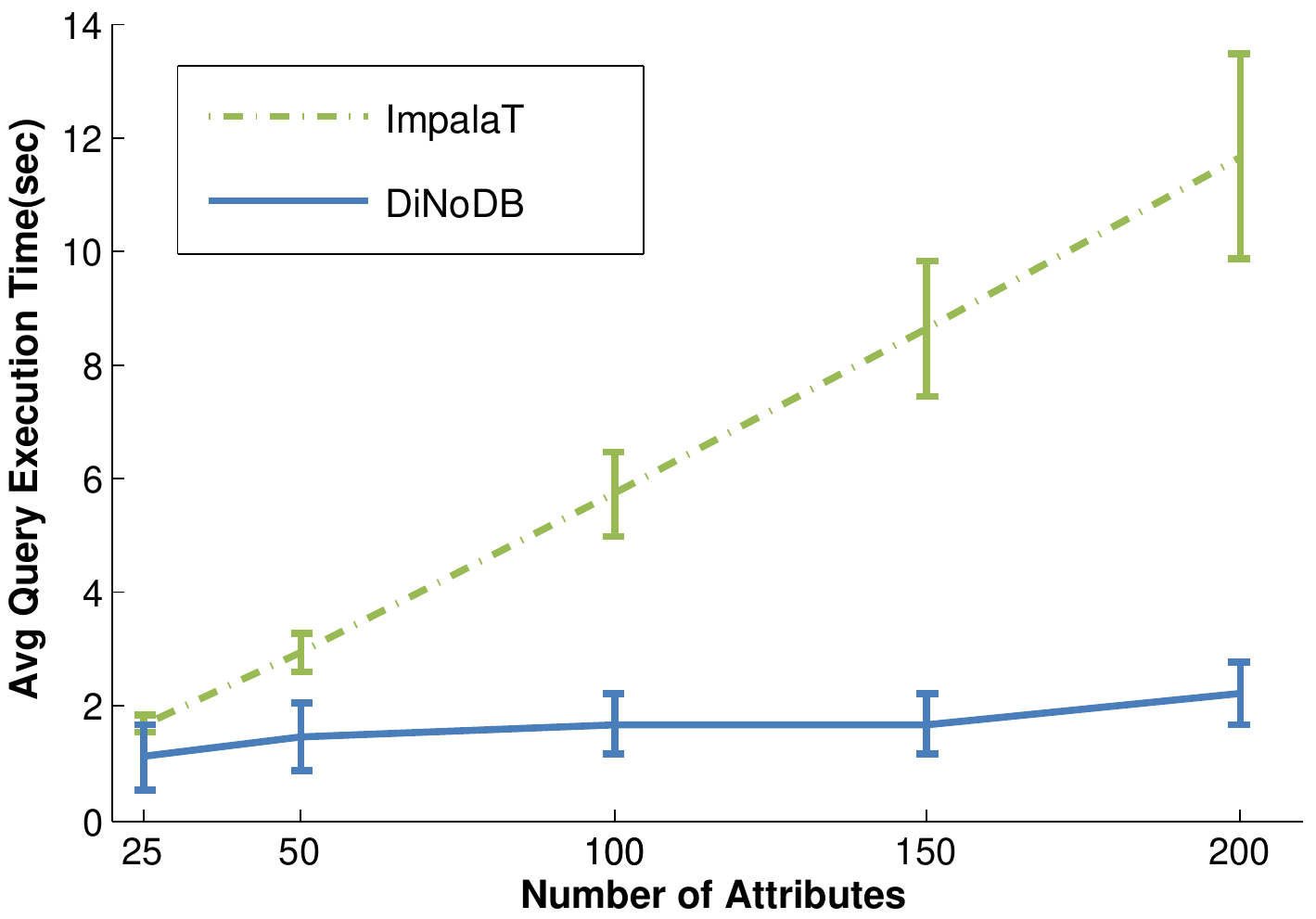}\label{fig:numattrscale}}
\subfigure[scaling the dataset size]{\includegraphics[width=\linewidth]{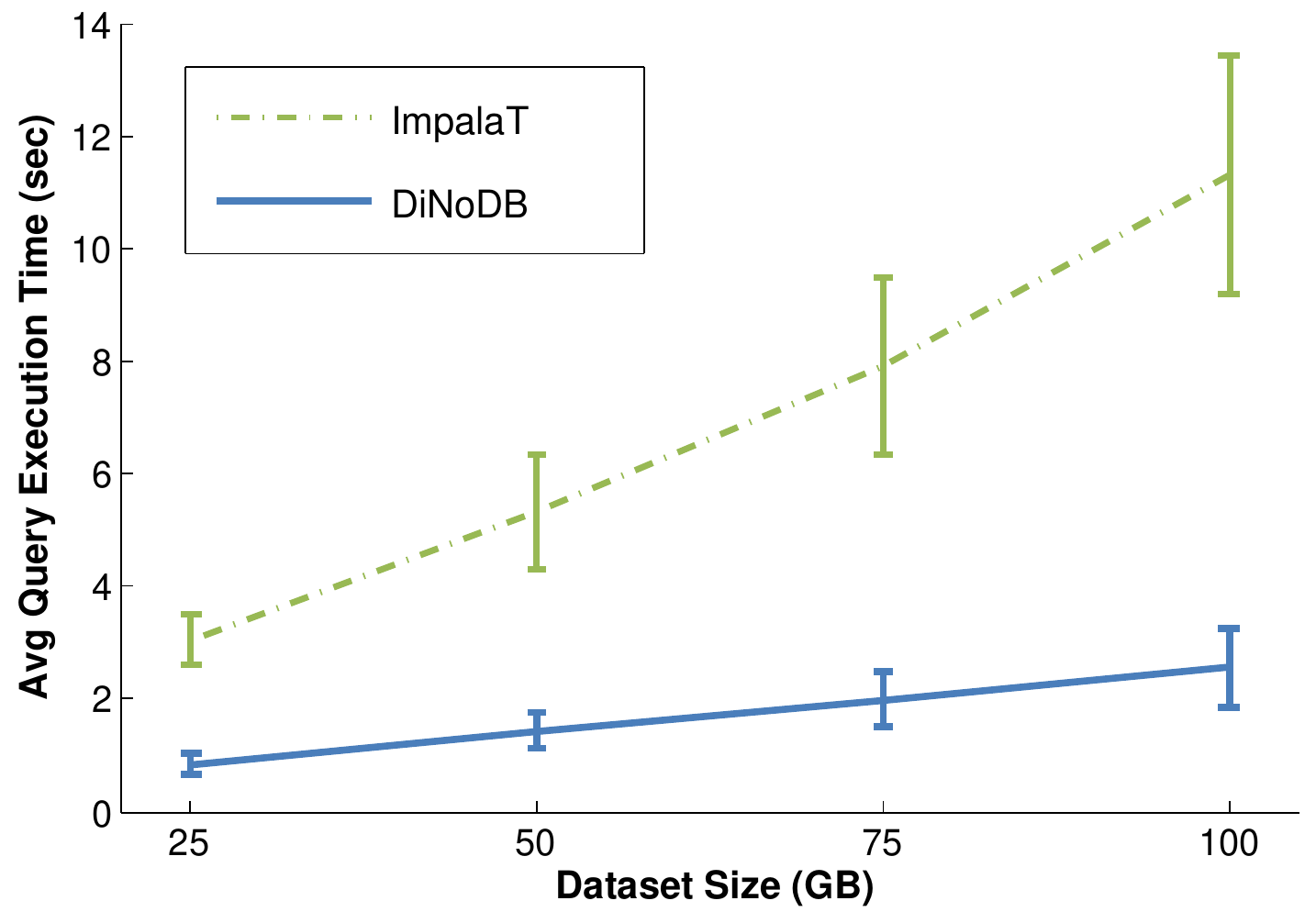}\label{fig:numrowscale}}
\caption{\db vs. ImpalaT: Scalability.}
\end{figure}

\textbf{Scaling the number of attributes.}
In this experiment, we fix the number of records (rows) of the synthetic dataset, but we vary the number of attributes in the range between 25 and 200 attributes (dataset size from 12 GB to 96 GB). For each of these datasets, we execute a sequence of 
50 SELECT PROJECT SQL queries in the same template as in Section~\ref{sec:randomqueries}. 
The average query execution time with standard deviation of \db and ImpalaT are shown in Figure~\ref{fig:numattrscale}.
\db average query execution time is almost constant when the number of attributes grows thanks to PM metadata; clearly, PM metadata size grows with the number of attributes in the data.
Instead, ImpalaT scales linearly with the number of attributes since it needs to parse every attribute (or byte) in a row: hence, it needs more CPU cycles when there are more attributes. 
ImpalaT has roughly the same performance as \db when there are 25 attributes. 
However, if the number of attributes exceeds 50, 
which is often the case the data analytics scenarios of our use cases, 
ImpalaT needs much more time to execute a query than \db.

\textbf{Scaling the dataset size.}
In this experiment, we keep constant the number of attributes to 100, and vary the number of records (rows) in the synthetic dataset. 
We compare \db and ImpalaT when dataset size ranges between 25 and 100 GB, 
by executing 50 SELECT PROJECT SQL queries on each dataset.
The average query execution time with standard deviation is shown in Figure~\ref{fig:numrowscale}. 
We observe that both ImpalaT and \db average query latency scale linearly with the dataset size. 
However, the slope for \db is less steep than that of ImpalaT.

\textbf{Discussion.}
The \plugin that generate PMs are crucial for the performance of \db, 
especially for datasets with many attributes. 
Note, that our prototype implementation of \db does not enjoy 
many of the important optimizations that are available for Impala, 
such as efficient data type handlers, just-in-time compilation and so on.

\subsection{Experiments with real life data}
\label{sec:real}
In this section, we give two examples of use cases we describe in Section~\ref{sec:usecases} using real life data. 
One is a topic modeling use case and the other is a data exploration use case.
In both examples, we show the query performance in different systems 
as well as the overhead of \plugin in the batch processing phase.

\subsubsection{Experiment on machine learning}
\label{sec:real_ma}

Here, we focus on a topic modeling use case, described in Section~\ref{sec:ml}. 
We use a 40 GB dataset collected by Symantec\footnote{Symantec: www.symantec.com}, 
consisting of roughly 55 million emails (in JSON format) 
tagged as spam by their internal filtering mechanism. 
To better understand the features of these spam emails, 
data scientists in Symantec 
often try to discover topics that occur in these collections of emails. 
In this experiment, 
we first play the role of a data scientist involved in topic modeling, 
and use Apache Mahout (version 0.11), a scalable machine learning library for Hadoop MapReduce. 
The CVB algorithm is iterative, 
and thus consists of multiple Hadoop MapReduce jobs with many intermediate data outputs. 
Since users are usually only interested in the final output, 
we instruct our \plugin to generate metadata solely in the last stage, 
upon algorithm termination, in which the distribution of documents and topics is finalized.

\textbf{\plugin overhead.}
First, we compare the Mahout topic modeling job 
to the one assembled with our \plugin, 
to study the overhead that it may impose on the pre-processing phase.
In general, machine learning algorithms executed 
on large datasets take quite a long time to complete. 
Topic modeling is not an exception; in our experiments, 
we set the two most important parameters of CVB to 20 topics 
and 5 iterations\footnote{These parameters have been discussed 
with domain experts, who require a reasonable number 
of topics for manual inspection.}. 
We run both the Mahout topic modeling job with \plugin and without \plugin \footnote{Native Mahout outputs results of topic modeling in binary through \emph{SequenceFileOutputFormat} class and needs command \emph{vectordump} to transform data to text format. In contrast, in our experiment we let Mahout output results in text format directly by replacing SequenceFileOutputFormat class with \emph{TextOutputFormat} class (without \plugin) or \emph{DiNoDBTextOutputFormat} class (with \plugin)}.
The resulting output file is a text file of roughly 23 GB, containing the probability matrix with 55 million rows and 21 columns.
The metadata which is generated by our 
\plugin mechanism is about 880 MB (with 1/5 sampling rate for positional map).

\begin{figure}
\centerline{\includegraphics[width=1\linewidth]{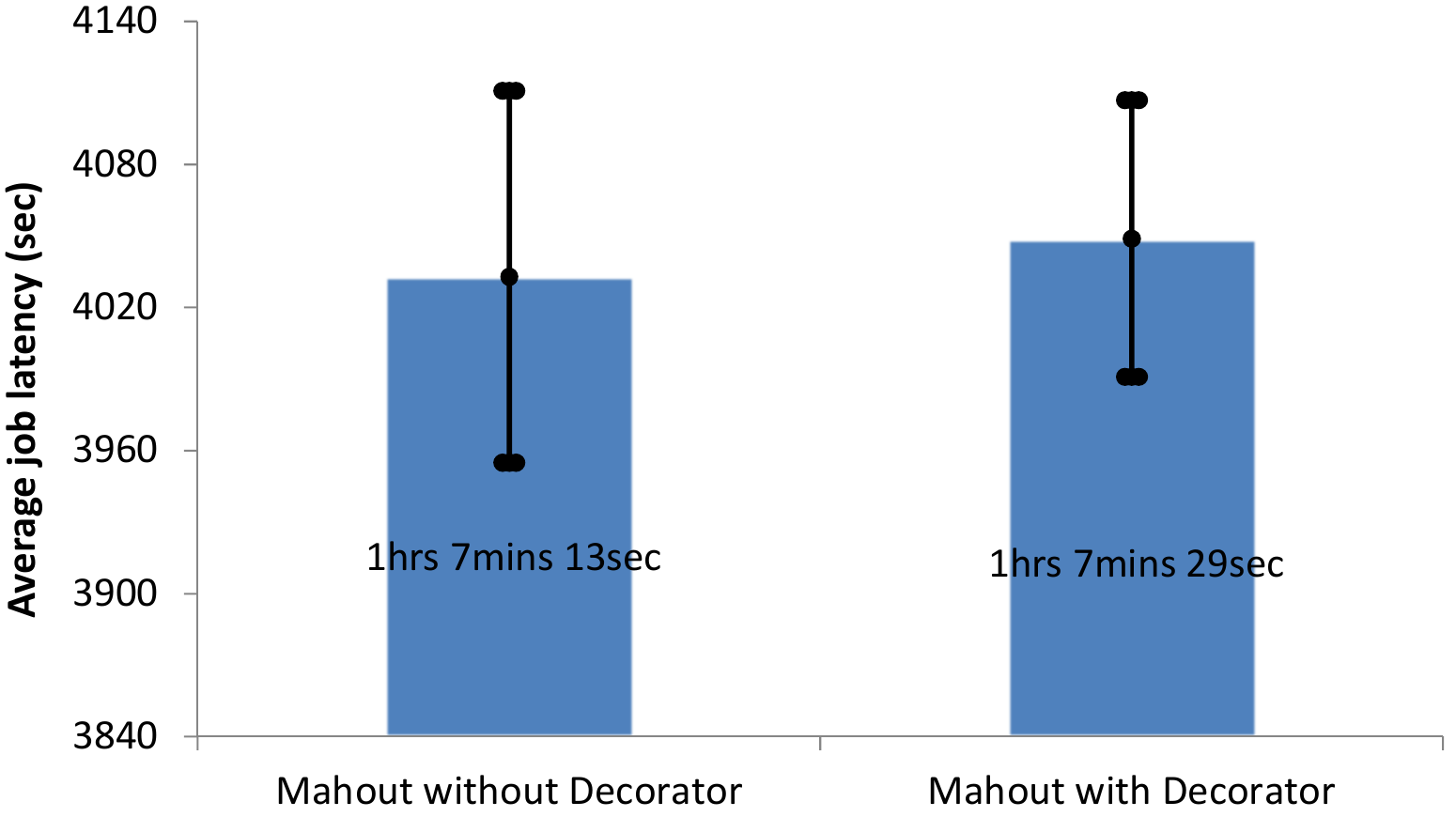}}
\caption{The processing time of the last stage of topic modeling with Mahout.}
\label{fig:overheadmahout}
\end{figure}

The overall runtime of Mahout topic modeling job is approximately 5 hours and 40 minutes.
In particular, we run the last stage
both without \plugin and with \plugin 10 times to measure the overhead of \plugin. The average processing time with standard deviation of the last stage is shown in Figure~\ref{fig:overheadmahout}, which shows that the overhead of \plugin is about 16 seconds.
This overhead, relative to the processing time of the last stage and the overall runtime of Mahout topic modeling job, demonstrates that the overhead introduced by our \plugin is essentially negligible.

\begin{figure}
\centerline{\includegraphics[width=1\linewidth]{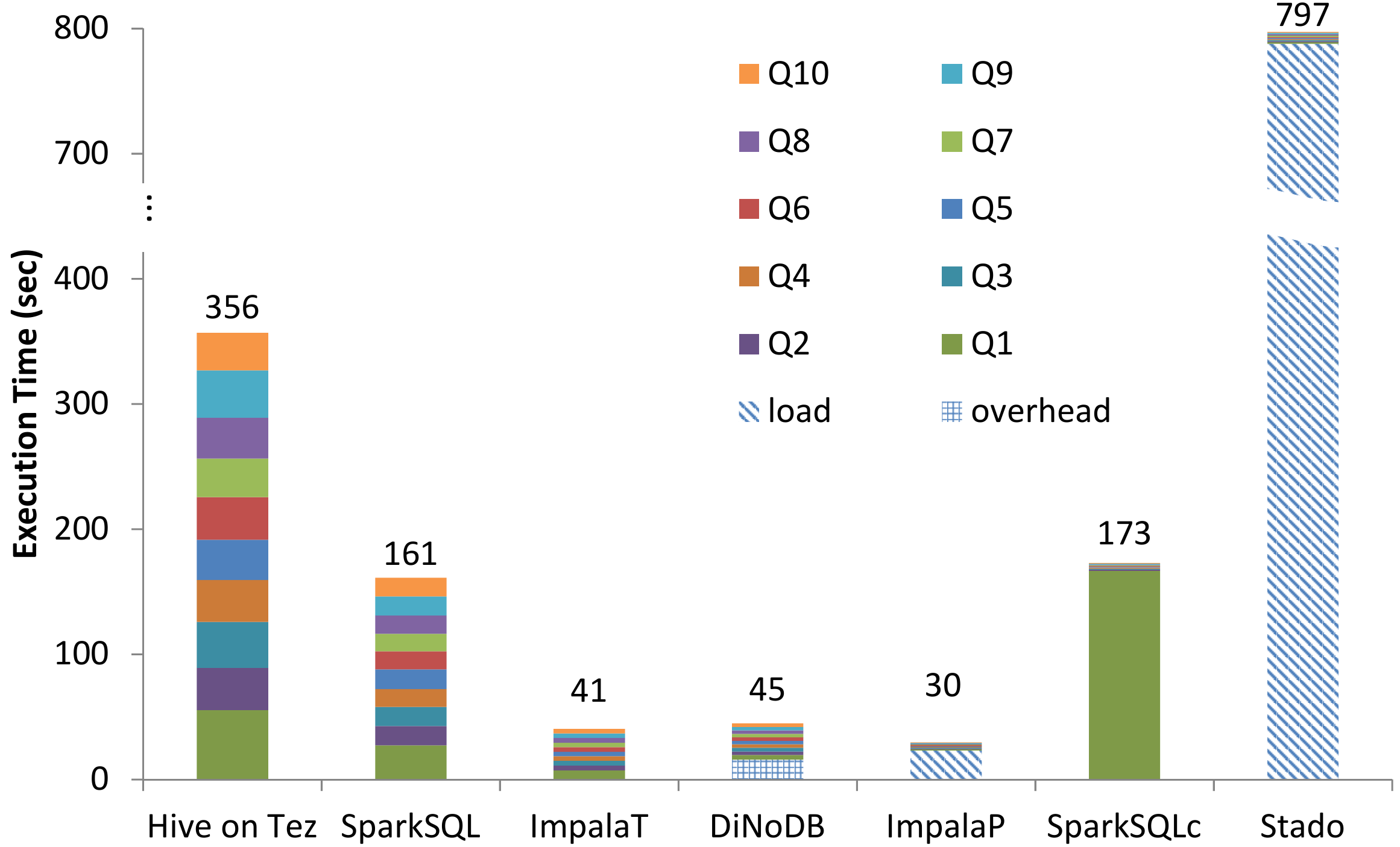}}
\caption{Query execution time in machine learning use case (Symantec dataset).}
\label{fig:mahoutcvbquery}
\end{figure}

\textbf{Query performance.}
In this experiment, we query the output of the topic-modeling phase:
the output data file is a ``doc-topic'' table, 
in which each row consists in a document identifier and the probability 
that such document belongs to each of the 20 topics. 
Hence, the output data file has one \texttt{INT} attribute (\texttt{docid}, 
which also serves as the key attribute) and 20 \texttt{FLOAT} attributes (probabilities). 
In order to understand if spam emails are well assigned to different topics, 
we would like to know which subset of emails have the highest probability to be in each topic. 
Therefore, the queries we execute with \db choose the top-10 spam emails per topic, sorted by probability measure:
\texttt{select docid, p\_topic\_x from table order by p\_topic\_x desc limit 10}, 
where \texttt{docid} is document (email) id and \texttt{p\_topic\_x} is the probability of that document belonging to topic\_x. 
Note that, since the queries are not based on the selectivity of the key attribute (\texttt{docid}), 
\db does not use VI, 
and solely relies on the PM 
file to improve performance. 

The result of a 10-query sequence for each system is shown in Figure~\ref{fig:mahoutcvbquery}. 
For a fair comparison, we penalize \db by adding the 16 seconds overhead of \plugin (as seen in Figure~\ref{fig:overheadmahout}) to the query execution time of DiNoDB, although we believe that users are much more sensitive to the latency in interactive analytics than the latency in batch processing.
With the help of positional map file generated by the \plugin, 
DiNoDB still achieves one of the shortest execution times for this sequence of queries, being only slightly slower only from ImpalaP and ImpalaT even though the overhead from batch processing was taken into account.
Notice that if the overhead of \plugin in the batch processing phase is not considered in the query execution time, \db achieves the shortest execution time.
Like in Section~\ref{sec:synthetic} SparkSQLc, ImpalaP and Stado have very short single query execution cost, 
which is less than 1 second, but they all need to first load data.
Note that, if we compare the loading time of SparkSQLc and ImpalaP with their loading time in section~\ref{sec:synthetic}, we find that Impala has a better support for data type \texttt{FLOAT} than SparkSQL.

\subsubsection{Experiment on data exploration}
\label{sec:real_de}

In this section, we focus on the data exploration use case, described in Section~\ref{sec:de}. We use a trace file which is the result of merging all the log files of Ubuntu One\footnote{\url{http://en.wikipedia.org/wiki/Ubuntu_One}} servers for 30 days (773 GB of CSV text). 
Now, let's assume that a user is interested in knowing the features of files stored in Ubuntu One. To better analyze this trace, this user filters out unnecessary information like server's RPC logs and creates a new data file called ``FileObject'' in which original trace is reorganized based on files. 
So, this user chooses to write Hadoop programs 
to pre-process the trace file.

\begin{figure}
\centerline{\includegraphics[width=1\linewidth]{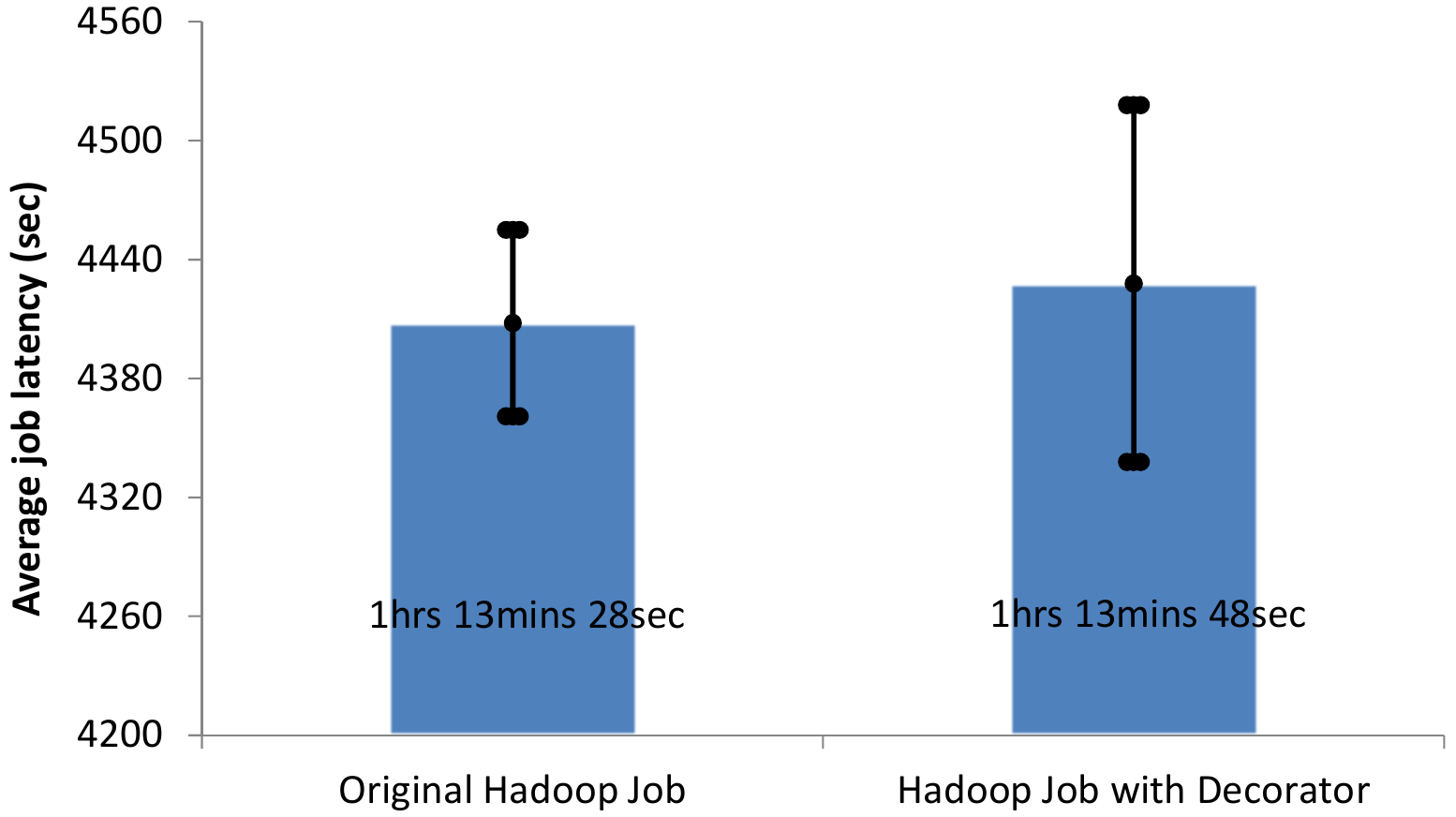}}
\caption{The batch processing time in data exploration use case
(Ubuntu One dataset)}
\label{fig:overheadubuntu1}
\end{figure}

\textbf{\plugin overhead.}
Here, we compare the normal Hadoop pre-processing job 
to the one assembled with \plugin, 
to study the overhead brought by \plugin. 
In ``FileObject'' file, each row/record represents a file stored in Ubuntu One server. 
Each row has 26 attributes, giving detailed information, e.g., mime type, creation time, file size, etc. 
To produce this ``FileObject'' file, 
we run both the original Hadoop job and the Hadoop job with \plugin 10 times. 
The resulting output is about 40 GB in size, containing information of 137 million unique files. 
A 2.2 GB PM 
metadata file is generated by \plugin mechanism (with 1/10 sampling rate). 
As shown in Figure~\ref{fig:overheadubuntu1},
the average processing time of original Hadoop job is 1 hour and 13 minutes and 28 seconds.
With \plugin, the Hadoop job needs about 20 seconds extra time, which is 0.45\% of the total batch processing 
cost. 
This overhead of \plugin is also negligible as in the machine learning use case.

\begin{figure}
\centerline{\includegraphics[width=1\linewidth]{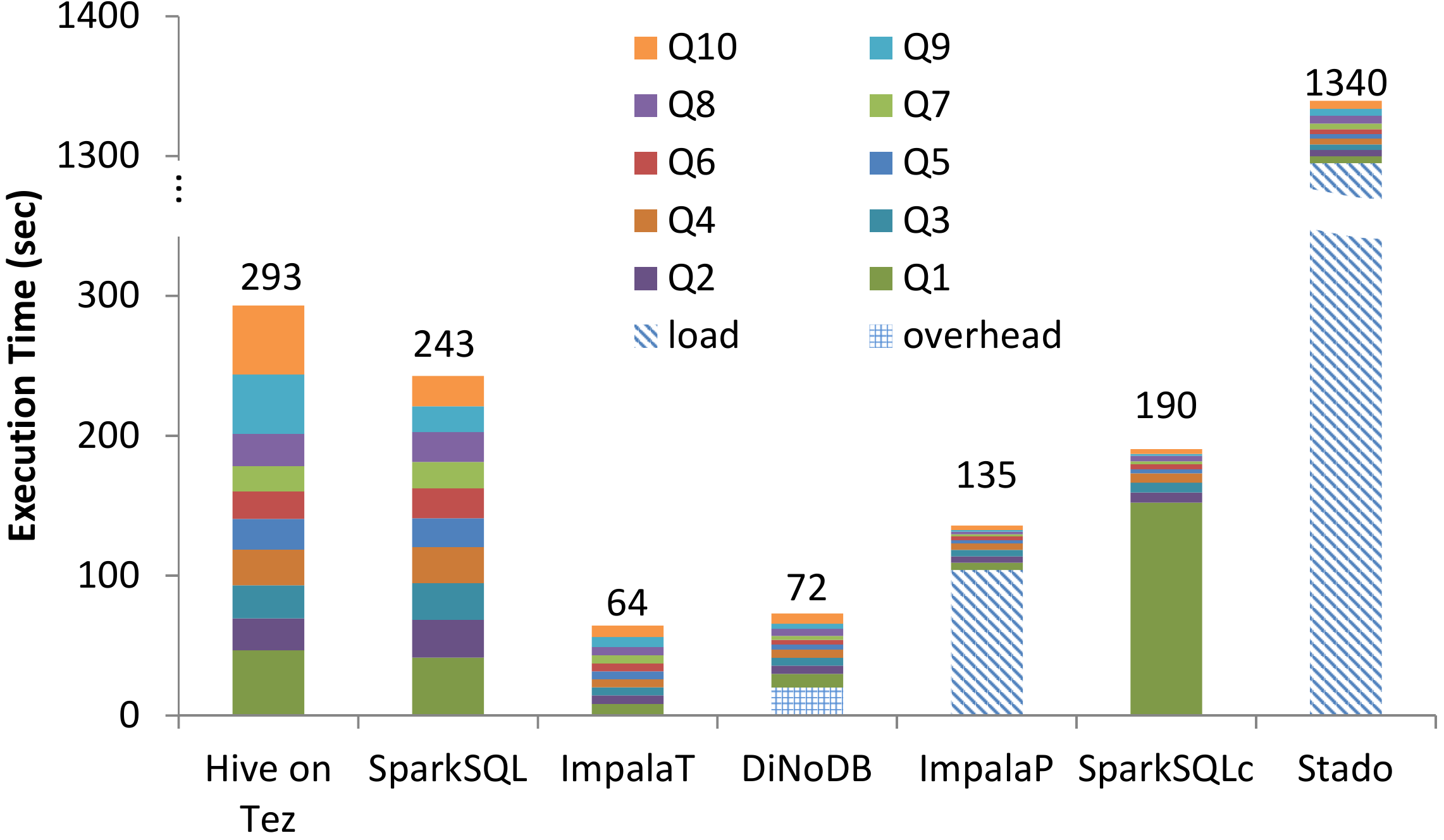}}
\caption{Query execution time in data exploration use case (Ubuntu One dataset).}
\label{fig:ubuntu1query}
\end{figure}

\textbf{Query performance.}
In this experiment, 
we compare the performance of the aforementioned systems when quering the result data file, ``FileObject''. 
The queries\footnote{\url{www.eurecom.fr/~tian/dinodb/ubuntuone.html} provides details on the dataset schema and the queries we used in our experiments.} we use in this experiment compute, for example, the number of distinct file extensions that users of the Ubuntu One service store, how many times the most popular file is downloaded, etc.
The result of a 10-query sequence is shown in Figure~\ref{fig:ubuntu1query}. 
For a fair comparison the 20 seconds overhead of \plugin (as seen in Figure~\ref{fig:overheadubuntu1}) is added to the query execution time of \db.
We find that query execution time of SparkSQLc, ImpalaP and Stado is much longer than in the previous experiments. That's because the queries executed in this experiment are more complex, by selecting more attributes and using operators like group by and aggregation. 
\db outperforms alternative systems except ImpalaT when considering total query execution time including \plugin overhead.

We note that, in this experiment as well as 
the respective experiment in Section~\ref{sec:real_ma}, 
ImpalaT query performance is on par with that of \db (without \plugin overhead). 
This is because when there are not many attributes (e.g., 20), the advantage of \db is not that obvious,
as we demonstrated in Section~\ref{sec:scalability}.

\subsection{Impala with \plugin}

Metadata generated by \plugin in the batch processing phase
can also be beneficial to other systems: they are not confined to be used in conjunction with \db. 
In this section, we present how Impala performs when we implant the \plugin in the workflow.

\begin{figure}
\centerline{\includegraphics[width=1\linewidth]{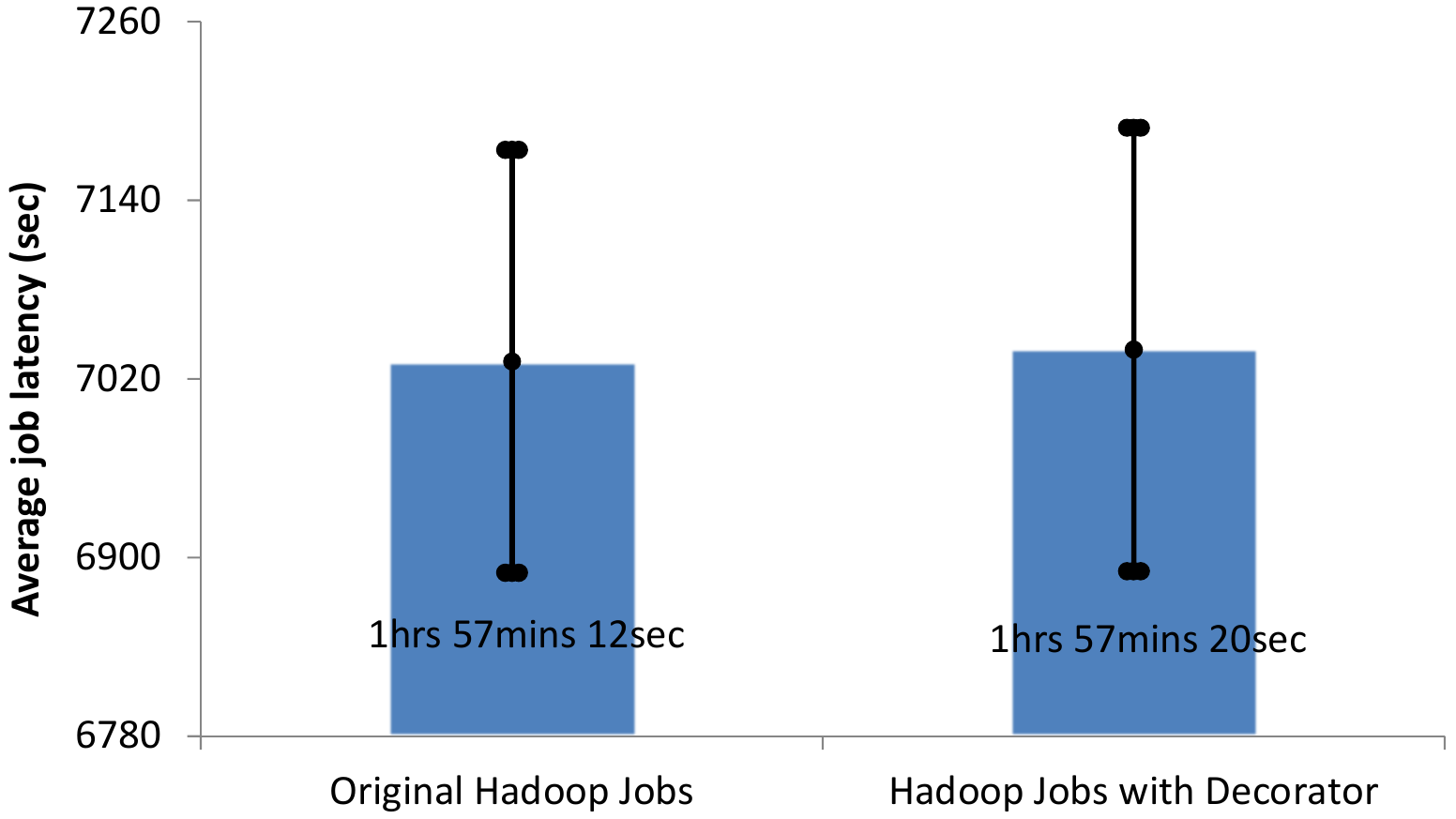}}
\caption{\plugin overhead to generate statistics.}
\label{fig:overheadubuntu1stat}
\end{figure}

In this experiment, an Impala user operates on the output data generated by a batch processing phase on the Ubuntu One server logs described previously. Let's assume the user wants to compute the number of times that files of a particular type, which are stored on the Ubuntu One service, are downloaded during a given period of time. 
In this case, in addition to the ``FileObject'' output file, a ``DownloadRecord'' output file is also produced in the batch processing phase. In ``DownloadRecord'' file, each record represents a download operation from Ubuntu One servers. Thus, our user instructs Impala to join two output data files.
We run Hadoop jobs with and without \plugin 10 times to measure the overhead of \plugin when only statistics decorator is used. The average batch processing time with standard deviation is shown in Figure~\ref{fig:overheadubuntu1stat} which shows that, on average, statistics decorator only brings an extra 8 seconds overhead.

\begin{figure}
\centerline{\includegraphics[width=1\linewidth]{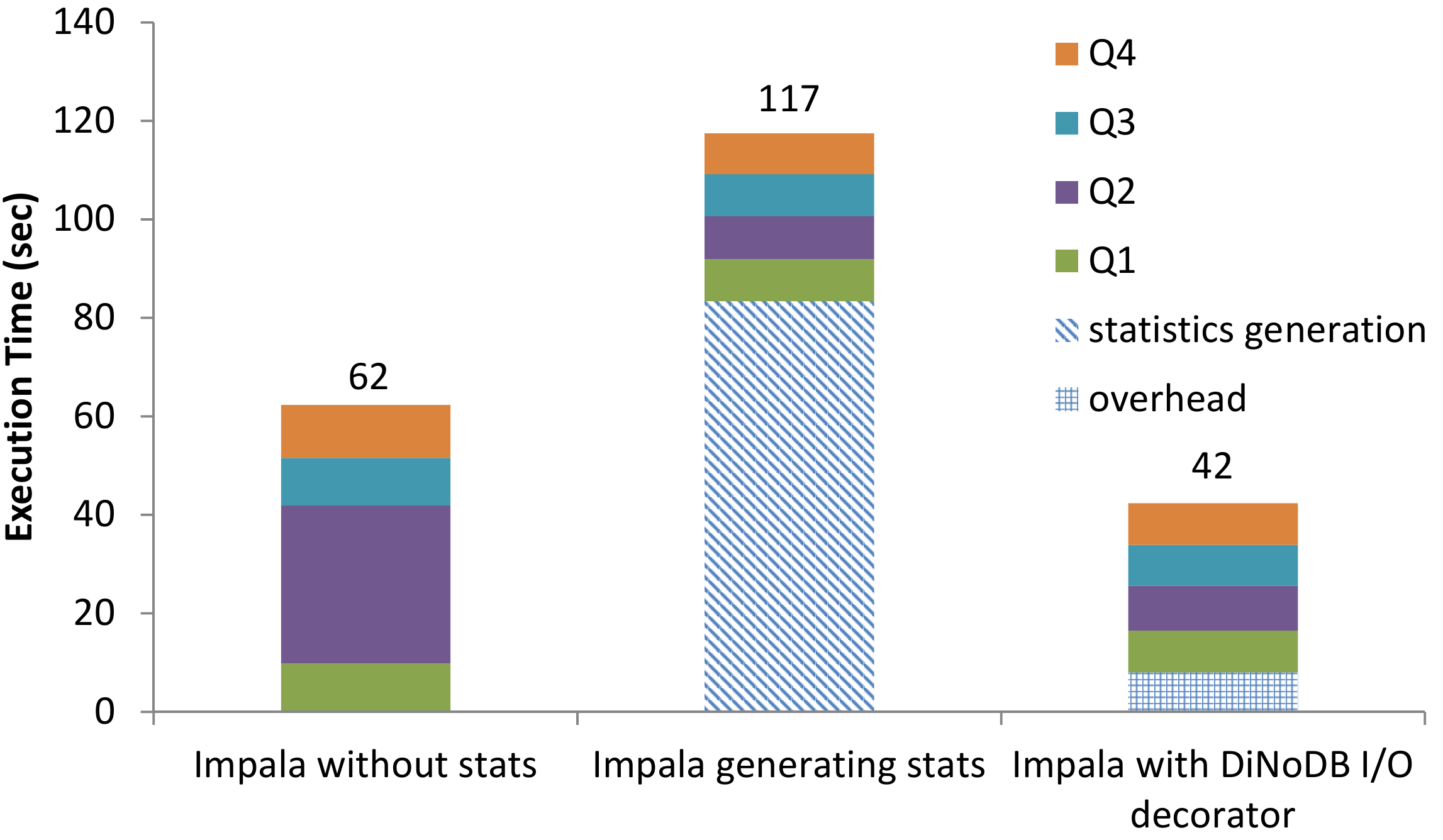}}
\caption{\plugin can be beneficial to other systems (e.g., Impala).}
\label{fig:impalajoin}
\end{figure}

Next, we compare Impala in three situations: i) executing queries without statistics, 
ii) executing queries after generating statistics using the built-in command ``Compute Statistics'' and 
iii) executing queries with statistics generated by \plugin during the 
batch processing phase\footnote{Statistics metadata can be passed to Impala by injecting them into the Impala metastore}. 
We execute 4 queries in each case and we show the results in Figure~\ref{fig:impalajoin} where the overhead of statistics decorator is also included.
In the first case, without statistics, Impala cannot choose an optimal query plan for the join operator, 
so the query execution time is the longest. 
In the second case, the query latency is quite short, 
but Impala needs to spend more than 1 minute to generate statistics before it can execute the queries. 
With \plugin, Impala achieves the lowest execution time even with the overhead from batch processing: this demonstrates the flexibility of our approach, and validates the design choice of piggybacking costly operations in the batch processing phase.

%% file: 05-related.tex
\section{Related work}
\label{sec:relwork}

Several research works and commercial products complement the batch processing nature of Hadoop/MapReduce~\cite{hadoop,Dean:2004:MSD:1251254.1251264} with systems to query large-scale data at interactive speed using a SQL-like interface. Examples of such systems include HadoopDB~\cite{Abouzeid:2009:HAH:1687627.1687731} and Vertica~\cite{Vertica}. These systems require data to be loaded before queries can be executed: in workloads for which data-to-query time matters, for example due to the ephemeral nature of the data at hand, the overheads due to the load phase, crucially impact query performance. In \cite{Abouzied:2013:ILA:2452376.2452377} the authors propose the concept of ``invisible loading'' for HadoopDB as a technique to reduce the data-to-query time; with invisible loading, the loading to the underlying DBMS happens progressively and on demand during the first time we need to access the data.
In contrast to such systems, \db avoids data loading and is tailored for querying raw data files leveraging 
metadata. Such files are built in \db with a lightweight piggybacking mechanism for workloads involving a preliminary data processing phase such as machine learning and data exploration use cases.

Shark~\cite{Xin:2013:SSR:2463676.2465288} presents an alternative design: it relies on a novel distributed shared memory abstraction called  Resilient Distributed Datasets (RDDs) \cite{Zaharia:2012:RDD:2228298.2228301} to perform most computations in memory while offering fine-grained fault tolerance. Shark builds on Hive~\cite{Hive} to translate SQL-like queries to execution plans running on the Spark system \cite{spark-project}, hence marrying batch and interactive data analysis.  Recently, SparkSQL \cite{SparkSQL} was announced as the Shark replacement in the Spark stack, as the new SQL engine for Spark designed from ground-up. The main characteristic of SparkSQL is that, just like Shark, it works with Spark's RDDs. Hence, both SparkSQL and Shark require a variant of data loading, to transform the raw HDFS data file and bring it into the RDD representation, in order to benefit fully from the Spark's in-memory low-latency processing. In contrast, \db achieves low latency while working on raw files, entirely avoiding data loading.

Impala~\cite{impala_cidr2015} is a state-of-the-art massively parallel processing (MPP) SQL query engine that runs in Hadoop. As such, Impala is probably the closest system to \db. However, while co-located with Hadoop, Impala does not leverage the possible synergy between batch processing of Hadoop and analytics power of a MPP SQL query engine. In this paper, we exactly propose such a synergy, and build \db to validate our approach. Namely, in \db
batch processing generates metadata 
(e.g., positional maps and vertical indexes) that helps  expedite SQL analytical queries. 
In this work, we demonstrate that the synergy between batch processing and query engines is beneficial for the analytical phase.

SCANRAW\cite{Cheng:2014:PID:2588555.2593673,1795} is proposed as a novel database physical operator, which loads data speculatively using available I/O bandwidth.
\nodb~\cite{Alagiannis:2012:NEQ:2213836.2213864} is a centralized DBMS that avoids data loading and transformation prior to queries. \db leverages \nodb as a building block to obtain \db nodes, which are to be seen as enhanced version of \nodb (see Section~\ref{sec:dinodbnode} for detailed comparison between \db nodes and PostgresRaw). 
A critical difference between \db and these works is that, \db is a distributed, massively parallel system for large-scale data analytics integrated with Hadoop batch processing framework, whereas \nodb and SCANRAW are centralized database solutions.
Dremel~\cite{36632} and GLADE~\cite{Cheng:2012:GBD:2213836.2213936} adopt multi-level aggregation to overcome single node bottleneck. This approach could be applied to \db to improve its scalability in future work.

Finally, \db shares some similarities with several research work that focuses on improving Hadoop performance. For example, Hadoop++~\cite{Dittrich:2010:HMY:1920841.1920908} modifies the data format to include a Trojan Index so that it can avoid full file sequential scan. 
Furthermore, CoHadoop~\cite{Eltabakh:2011:CFD:2002938.2002943} co-locates related data files in the same set of nodes so that a future join task can be done locally without transferring data in network. 
However, these systems require users to write specific Hadoop program, which could be complex and hard. Therefore, neither Hadoop++ nor CoHadoop are suitable for interactive raw data analytics, like \db is.

To conclude, we compare our current version of \db with a preliminary version, presented in \cite{DiNoDB:Data4U}. The preliminary version of \db used HadoopDB to orchestrate \db nodes, 
which suffered from the inherent overhead of the HadoopDB framework when it came to interactive queries.
 Therefore, we replaced HadoopDB with a massively parallel architecture, but had to address fault tolerance of such an architecture. Moreover, 
our current version of \db introduces \plugin, which are easy to implement and help couple batch layer and serving layer more seamlessly.
As a result, our current version of \db 
outperforms state-of-the-art analytics systems such as SparkSQL and Impala, for temporary data.

%% file: 07-conclusions.tex
\section{Conclusion}
\label{sec:conclusion}

Parallel data processing systems such as Hadoop MapReduce have received increasing attention, for their promise to analyze any amount or kind of data. However, with such systems, users had to move and load the data produced in the batch-processing phase into a fast relational database, to achieve interactive-speed data manipulation. 
The specter of ``leaving something important behind'' related to data movement and adaptation has led academics and the industry to designing new systems that would expose a standard, interactive way to manipulate data, while being fully integrated in a growing ecosystem of data processing tools.

In this work, we presented the architecture of \db, a distributed system tuned for interactive-speed queries on temporary data files generated by large-scale batch-processing frameworks. As shown by our extensive experimental evaluation, for the use-cases \db targets -- ad hoc queries on a narrow processing window, our system outperforms current SQL-on-Hadoop solutions. \db uses a decorator mechanism that enhances the standard Hadoop I/O API and piggybacks the creation of auxiliary metadata required for interactive-speed query performance. In addition, \plugin seamlessly integrate with existing frameworks and distributed storage systems.

Our experimental evaluation, that we do on both synthetic and real-world datasets, highlights the key benefits of \db in a number of prominent use cases, making it suitable for a wide range of ad-hoc analytical workloads.